\documentstyle[12pt,A4]{article}

\input epsf
\epsfverbosetrue
\begin{document}

\renewcommand{\thefootnote}{\alph{footnote}}
\begin{titlepage}
\begin{tabbing}
\hspace{11cm} \= HIP -- 1997 -- 22 / TH \\
\> LTH -- 394\\
\> \today
\end{tabbing}
\begin{centering}
\vfill
{\Large\bf Flux-tube structure and $\beta$-functions in SU(2)\\[1ex]}
\vspace{1cm}

P.~Pennanen\footnotemark[1],
A.M.~Green\footnotemark[2]$^,$\footnotemark[3],
\vspace{0.25cm}

Helsinki Institute of Physics\\ $\,^b$Department of Physics\\ 
University of Helsinki, Finland\\
\vspace{0.5cm}

and C.~Michael\footnotemark[4]\\
\vspace{0.25cm}

Theoretical Physics Division, Dept. of Math. Sciences,
University of Liverpool, Liverpool, UK.

\end{centering}
\setcounter{footnote}{1}
\footnotetext{email: {\tt petrus.pennanen@helsinki.fi}}
\setcounter{footnote}{3}
\footnotetext{email: {\tt green@phcu.helsinki.fi}}
\setcounter{footnote}{4}
 \footnotetext{email: {\tt cmi@liv.ac.uk}}
 \setcounter{footnote}{0}
\renewcommand{\thefootnote}{\arabic{footnote}}
\vspace{1.5cm}

\begin{abstract}  The spatial distribution of the action and 
energy in the colour fields of flux-tubes is studied in lattice $SU(2)$ field
theory for static quarks at separations up to 1 fm at $\beta=2.4,\,2.5$. 
The ground and excited states of the colour fields are 
considered. Sum rules are used to get estimates of generalised 
$\beta$-functions. 

\end{abstract}
\vfill

\end{titlepage}
\section{Introduction}

Non-perturbative phenomena of QCD such as confinement can be explored
using  Monte Carlo simulations of lattice gauge theory. The potential
$V(R)$ between  two static quarks at separation $R$ in quenched QCD is
one of the simplest manifestations of confinement and has been studied
intensively. At large $R$  the potential rises linearly as predicted by
the hadronic string model. One  can also measure the spatial
distribution of the colour fields around such  static quarks in order to
get a detailed picture of the confining flux tube.  In Ref.
\cite{gre:96}, which contains references to earlier work,  this was
done for the  ground state and the first excited
state of the two-quark potential, having the symmetries of
the $A_{1g}$ and  $E_u$ representations, respectively, of the lattice
symmetry group $D_{4h}$. Transverse and longitudinal profiles of
chromoelectric and -magnetic fields  were compared with vibrating string
and dual QCD models for the flux  tube, with the latter model
reproducing quite well the shape of the energy profile  measured on a
lattice. Instead of SU(3), the gauge group used was SU(2), which  is
more manageable with present-day computer resources and is expected to
have very similar features of confinement. This is reflected in the fact
that the flux-tube models considered do not distinguish between SU(2)
and SU(3). 

The method used to study the colour fields on a lattice is to measure 
the correlation of a plaquette $\Box\equiv{1 \over 2} {\rm Tr}
(1-U_{\Box})$ with the Wilson loop $W(R,T)$ that represents the static
quark and antiquark  at separation $R$.  When the plaquette is located
at $t=T/2$ in the   $\mu,\ \nu$ plane, the following expression
isolates, in the limit  $T \to \infty$, the contribution of the colour
field at position ${\bf r}$:
 \begin{equation}  \label{fmnT}
 f_R^{\mu \nu}({\bf r})=\left[{\langle
W(R,T)
  \Box^{\mu \nu}_{\bf r}\rangle}
-\langle W(R,T)\rangle \langle \Box^{\mu \nu}\rangle
\over {\langle W(R,T)\rangle} \right].
\end{equation}

In the naive continuum limit these contributions are related to the mean
squared fluctuation of the Minkowski colour fields by
 \begin{equation}
\label{fmn}
 f_R^{ij}({\bf r})\rightarrow  \frac{a^4}{\beta}B^2_k({\bf r}) \quad
{\rm with} \ i,\ j,\ k\   {\rm cyclic \ \ \ and} \quad
f_R^{i4}({\bf r})\rightarrow -\frac{a^4}{\beta}E^2_i({\bf r}).
\end{equation}
When the interquark separation axis is chosen as the 1-axis the 
squares of the longitudinal 
and transverse electric and magnetic fields can be identified as
 \begin{equation} \label{ae3}
 {\cal E}_L =  f^{41},\; {\cal E}_T =  f^{42,43},\; {\cal B}_T =
f^{12,13},\; {\cal B}_L =   f^{23,32}.
 \end{equation}
These can then be combined naively to give the action density
 \begin{equation} \label{ae}
S({\bf r})=-({\cal E}_L + 2  {\cal E}_T  + 2  {\cal B}_T + {\cal B}_L )
 \end{equation}
 \noindent and the energy density
\begin{equation}
 \label{ae2}
 E({\bf r})=E_L({\bf r})+2E_T({\bf r})=
-({\cal E}_L - {\cal B}_L )- 2 ( {\cal E}_T  -  {\cal B}_T )
 \end{equation}
of the gluon field. 

Since in this work we use a plaquette to probe the colour  flux,
the spatial size of the probe will decrease as the lattice spacing $a
\to 0$. To define a continuum limit of the colour flux distributions,
one would have to use a probe of a fixed physical size as $a \to 0$. In
this work we wish to compare flux distributions at different lattice
spacing. One special tool that is available, when a plaquette is used to
probe  the colour flux with the Wilson gauge action, is that exact
identities can be  derived for the integrals over all space of the flux
distributions.  These   sum rules~\cite{mic:87,mic:96} relate spatial
sums  of the colour fields measured using Eq.~\ref{fmnT} to the static
potential $V(R)$ via generalised $\beta$-functions, which show
how the bare  couplings of the theory vary with the generalised lattice
spacings $a_\mu$ in four  directions. One can think of these sum rules
as providing the  appropriate anomalous dimension for the colour flux
sums.  This normalises the colour flux and provides a guide for
comparing  colour flux distributions measured at different $a$-values. 
The full  set of sum rules~\cite{mic:96} allow these generalised
$\beta$-functions  to be determined at just one $\beta$-value~\cite[and
references therein]{mic:96b}.  Here we investigate this further by
comparing estimates at two  different $\beta$-values. This can also help
to clear up some inconsistencies  between  the $\beta$-function determination
from the sum rules at one $\beta$-value ~\cite{mic:96b} and  other
methods \cite{eng:95,pen:96b}.

In Ref. \cite{gre:96}  the simulations were  carried out at $\beta=2.4$
with a $16^3\times 32$ lattice. Here results from  similar computations
at $\beta=2.5$ with a $24^3\times 32$ lattice are  reported. Most of the
simulation and analysis techniques, such as use of a  variational basis
with different fuzzing levels, are the same and can be found  from Ref.
\cite{gre:96}. In addition to more accurate measurements of  flux-tube
profiles, we also present  estimates of $\beta$-functions  at
both $\beta=2.4$ and 2.5.

In Sect. 2 static quark-quark potentials $V(R)$ are extracted and in
Sect. 3 the corresponding flux tube profiles calculated. In Sect. 4,
$V(R)$ and the profiles are related by sum rules and various estimates
of the $\beta$-function are made. Some conclusions are made in Sect. 5.

%**************************
\section{Static potentials \label{spot}}
%**************************

We construct lattice operators to create and destroy
states  with  two static quarks at separation $R$ joined by  gluonic
paths which represent the  colour flux. The techniques we use to make
efficient operators with a large overlap with the ground state are
described in detail in Ref.~\cite{gre:96}. In order to improve the
signal we applied ``fuzzing'', where each spatial link is replaced by a
weighted sum of itself and its surrounding spatial staples, before the
correlations were measured. To investigate gluonic  excitations and
minimize their contribution to the ground state signal we  need a
variational basis, which was obtained by performing the measurements on
lattices with different levels of fuzzing.

At $\beta=2.4,\, 2.5$ fuzzing levels 40, 16, 0 and 40, 13, 2, 
respectively, formed the variational basis in the case of paths with
$A_{1g}$ symmetry. A three-state basis may be expected to give a
reasonable signal for the $A_{1g}'$ excitation by reducing  the
contamination from higher excitations with this symmetry.  For paths
with $E_u$  symmetry the fuzzing levels 16, 13 were used for the two
$\beta$'s with two  different transverse extents of the paths forming a
variational basis. These transverse extents were one and two lattice
spacings for small longitudinal lengths and  one and four lattice
spacings for larger $R$'s. In order to set the scale, the  lattice steps
$a(2.4)$ and $a(2.5)$ were determined  by fitting the two-body
parameterisation  
 \begin{equation} \label{VV0} V(R) = -
\left[\frac{e}{R}\right]_L + b_S R + V_0 \label{ev}  
 \end{equation} 
 to measured potentials at interquark separations  $R=2,3,4,6,8$,
$R=2,3,4,6,12$ at $\beta=2.4,\,2.5$ respectively. Here
$\left[\frac{1}{R}\right]_L$ is the latticized form of the Coulomb
potential $1/r$ due to one-gluon exchange.  The above range of $R$ was
chosen to  correspond to similar physical distance ranges. The usual
method for estimating the lattice spacing is to equate the dimensionless
value of $b_S$ from the  fit of Eq.~\ref{ev} to an experimental value.
This is equivalent to  utilizing $V(R)$ in the limit $R\rightarrow
\infty$. However, our experimental knowledge of the two-quark potential
comes from heavy mesons with r.m.s. radii around 1 fm.  An alternative
method due to Sommer \cite{som:94} compares the force from  $V(R)$ to
experimental values at a distance range more appropriate to these
mesons, i.e.  $r\approx 0.5$ fm. In practice the equation  $(R_0)^2
F(R_0) = c$ is used to find $R_0$, where $F(r)$ is the  force between
two static quarks at separation $r$. Various nonrelativistic  continuum
potential models give $aR_0\approx 0.49$ fm for $c=1.65$ and 
$aR_0\approx 0.66$ fm for $c=2.44$ \cite{pen:96b}. The resulting scales
and  ratios of lattice spacings $\rho\equiv a(2.4)/a(2.5)$ are shown in
Table \ref{tscale}. The values of $\rho$ from $b_S$ and Sommer's scheme
are seen to agree.

\begin{table}[htb]
\begin{center}
\begin{tabular}{l|c|c|c} 
$\beta$     & $c=1.65$    & $c=2.44$  & $\sqrt{b_S}=440$ MeV    \\ \hline
2.4         &  0.1098(5)  & 0.1183(5) & 0.1190(5)  \\
2.5         &  0.0778(4)  & 0.0839(4) & 0.0846(4)  \\ \hline
$\rho\equiv a(2.4)/a(2.5)$ & 1.412(14) & 1.410(13) & 1.406(11)
\end{tabular}
\caption{Lattice spacing $a$ determined using different methods}
\label{tscale}
\end{center}
\end{table}

The $\beta=2.5$  potentials above have the accurate  interpolations 
 \begin{eqnarray}
\label{VA1g}
 V(R)_{A_{1g}} &=& 0.555 + 0.0343 R - 0.280/R \;\\
\label{VEu}
  V(R)_{E_u} - V(R)_{A_{1g}} &=& 3.8/R -12.6/R^2 +24.6/R^3- 18.8/R^4 \;\\
 \label{VA1gp}
  V(R)_{A'_{1g}} - V(R)_{A_{1g}} &=& 5.94/R -23.6/R^2
+49.2/R^3-38.8/R^4\;
 \end{eqnarray}
The relations are valid for $2 \le R \le 12$, with no physical interpretation
intended for the $V(R)_{E_u}$ and $V(R)_{A'_{1g}}$ expressions. The $\chi^2$
values per d.o.f. are 0.09, 1.35 and 0.22 for the $A_{1g}$, $E_u$ and 
$A_{1g}'$ states respectively. The corresponding potential fits for 
$\beta=2.4$ are given in Ref. \cite{gre:96}.

Our variational basis is constructed from fuzzed link operators that  
represent creation or annihilation of two 
quarks at separation $R$, with the colour field in a specific state of
lattice symmetry.
These operators can be expanded in terms of the eigenstates of the transfer
matrix
\begin{equation}
|R\rangle=c_0|V_0\rangle+c_1|V_1\rangle+\dots
\end{equation}
with the measured correlation of a generalised Wilson loop given by
\begin{equation}
W(R,T)=\langle R_0|R_T\rangle =
c_0^2 e^{-V_0 T} \left(1 + h(T/2)^2 + \dots \right) \label{ewcorr}
\end{equation}
where
\begin{equation}
h(t)={ c_1 \over c_0} e^{-(V_1-V_0)t}\ .
\end{equation}
To minimize excited state contamination we need $h \ll 1$. 

As plaquettes in the middle of the generalised Wilson loop in
the time  direction are used to probe the colour fields, the relevant
estimate of  contamination is taken at $t=T/2$.  The measured
correlation is $\langle R_0| \Box_t |R_T\rangle$ instead of
Eq.~\ref{ewcorr}. This produces off-diagonal terms such as  $\langle
V_1|\Box_t|V_0 \rangle$ which increase the coefficient of excited state
contribution from $h$ to $2h$ for $T=2$, $t=1$. From the generalised
Wilson loop ratios at each $R$-value, we define  $V(T)= -\ln
[W(T)/W(T-1)]$  since its rate of approach to a plateau as $T \to
\infty$ enables us to estimate the excited state contamination to the
ground state.  We calculate $h$ from
 \begin{equation} |h(t=T/2)| \approx \frac{\lambda}{\lambda-1}
\sqrt{V(T-1)-V(T)} =
\lambda\frac{V(T-1)-V(T\rightarrow\infty)}{\sqrt{V(T-1)-V(T)}}.
 \end{equation} 
 Here the $T\rightarrow\infty$ extrapolated potential is defined as  
 $$
V(T\rightarrow\infty)\equiv V(T)-\lambda\frac{V(T-1)-V(T)}{1-\lambda},\ 
 \lambda\equiv e^{-(V_1-V_0)}. 
 $$ 
 In practice $\lambda$ was calculated from potentials at $T=1$. Table
\ref{th} shows the excited state contamination for states of the
two-body potential at $\beta=2.5$. 
It is seen that the $A_{1g}$ states are reasonably pure with the $E_u$
and $A_{1g}'$ states containing increasing amounts of contamination.
Note that  the method used to extract the higher excited state
contamination $h$ to the  first excited $A_{1g}$ state  is unreliable
since ground state contributions can dominate in principle.

\begin{table}[htb]
 \caption{Excited state contamination at $\beta=2.5$ as measured by $h$
\label{th}}
 \begin{center}
\begin{tabular}{l|l|c|c|c}
$R$ & state     & t=1   & 2     & 3 \\ \hline
2   & $A_{1g}$  & 0.012 & 0.005 & -    \\
3   & $A_{1g}$  & 0.027 & 0.012 & 0.009 \\
4   & $A_{1g}$  & 0.036 & 0.016 & 0.008  \\
6   & $A_{1g}$  & 0.073 & 0.034 & 0.016 \\
12  & $A_{1g}$  & 0.162 & 0.075 & 0.006 \\
2   & $E_u$     & 0.282 & 0.152 & 0.105 \\
3   & $E_u$     & 0.253 & 0.131 & 0.105 \\
4   & $E_u$     & 0.250 & 0.126 & 0.076 \\
6   & $E_u$     & 0.255 & 0.123 & 0.045 \\
12  & $E_u$     & 0.427 & 0.263 & 0.189 \\ 
2   & $A_{1g}'$ & 0.201 & 0.080 & 0.076 \\
3   & $A_{1g}'$ & 0.228 & 0.098 & 0.070 \\
4   & $A_{1g}'$ & 0.261 & 0.116 & 0.122 \\
6   & $A_{1g}'$ & 0.333 & 0.155 & 0.157 \\
12  & $A_{1g}'$ & 0.518 & 0.241 & 0.204 \\ 
\end{tabular}
\end{center}
\end{table}

%************************************************
\section{Colour field distributions \label{scol}}
%************************************************

Before embarking on the extraction of $\beta$-functions using sum rules
as  presented in the next section, it is of interest to check further
the quality of the input data for the sum rules by testing its scaling
properties between $\beta=2.4$ and $2.5$. Essentially the  input data is
of two distinct forms -- the two quark potential $V$ and the  spatial
sums of the colour fields. In the past the scaling properties of $V$
have been confirmed many times and will not be repeated here. However,
scaling of the different colour field combinations is less clear, since 
the colour fields are measured using a plaquette, whose physical size
changes with $\beta$. This is relevant because only observables with the
same  physical size at different values of the coupling have a continuum
limit.  In the  case of the  three-dimensional sums  over the
colour fields, the lattice sum rules  provide the appropriate
normalisation as $a \to 0$.  Even in this case, the scaling behaviour is
only known  after the divergent self-energies are subtracted. However,
other observables,  like the two-dimensional sums over transverse planes
or the transverse profiles of the flux-tube, do not have a
well defined scaling behaviour, but it is  still of interest to explore
how similar the profiles are at the two values of  coupling used.  

 From the relations in Eq.~\ref{fmn} and  sum-rules to be
presented in Eqs. \ref{TASU}--\ref{TEPSU}, we can see that the  measured
action sums must be  multiplied by the anomalous dimension $b/\beta$ to
get the physical value, while the energy sums have a correction $f$
which goes to one in the  continuum limit.  Here $b,\, f$ are
generalised $\beta$-functions  to be discussed later. As will be seen in
sect. \ref{sb} below, the  differences in the values of $b$ and $f$
between $\beta=2.4$ and 2.5 are sufficiently small to neglect  them in
the following plots; e.g. $b(2.4)/b(2.5)=0.97(7)$. Thus the
normalisation  of the overall three dimensional sum over colour flux can
be treated as almost  constant in our $\beta$ range. In turn this
implies that we should compare more differential  distributions using
this scale.

\subsection{Spatial sums}

\begin{figure}[h]
\hspace{0cm}\epsfxsize=400pt\epsfbox{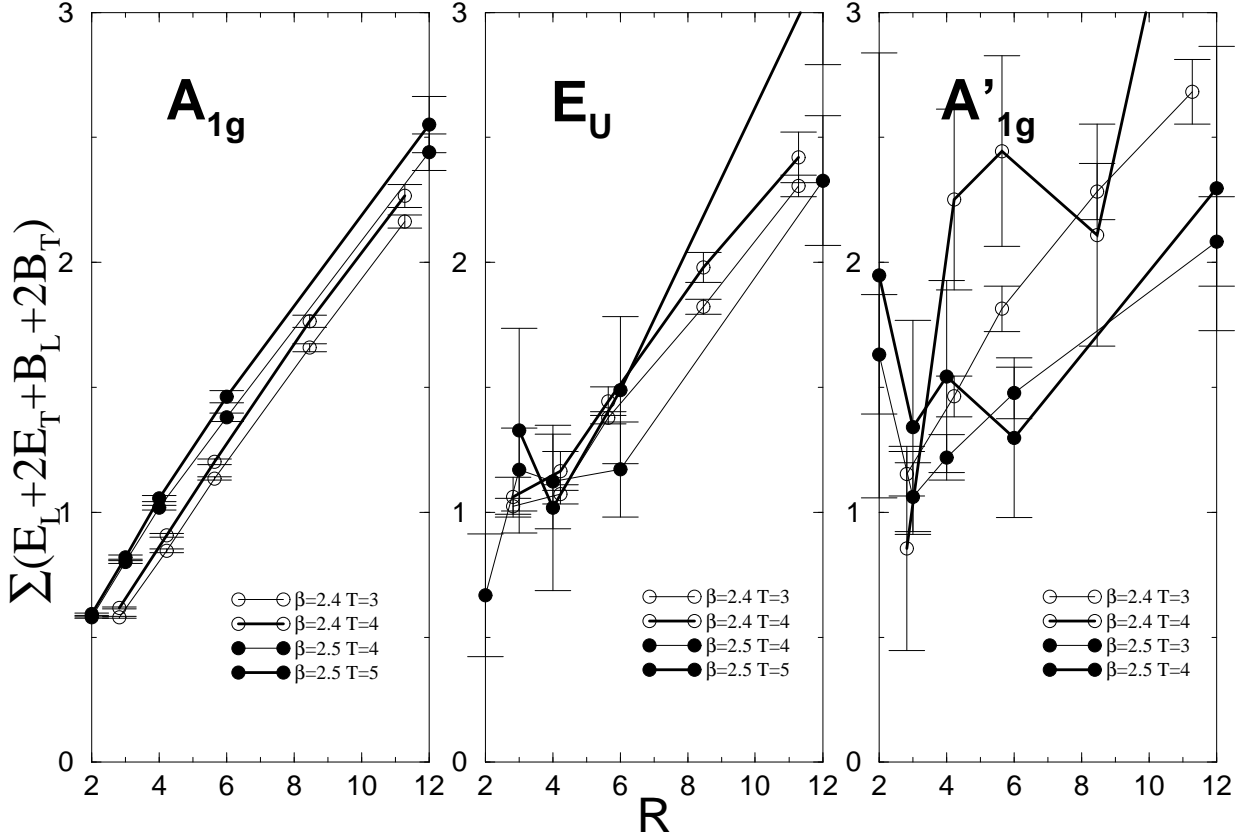}
 \caption{The scaling of action ($S$) summed over the spatial lattice
for paths with $A_{1g}$, $E_u$ and $A_{1g}'$ symmetries.}
 \label{facts}
\end{figure}

\begin{figure}[h]
\hspace{0cm}\epsfxsize=400pt\epsfbox{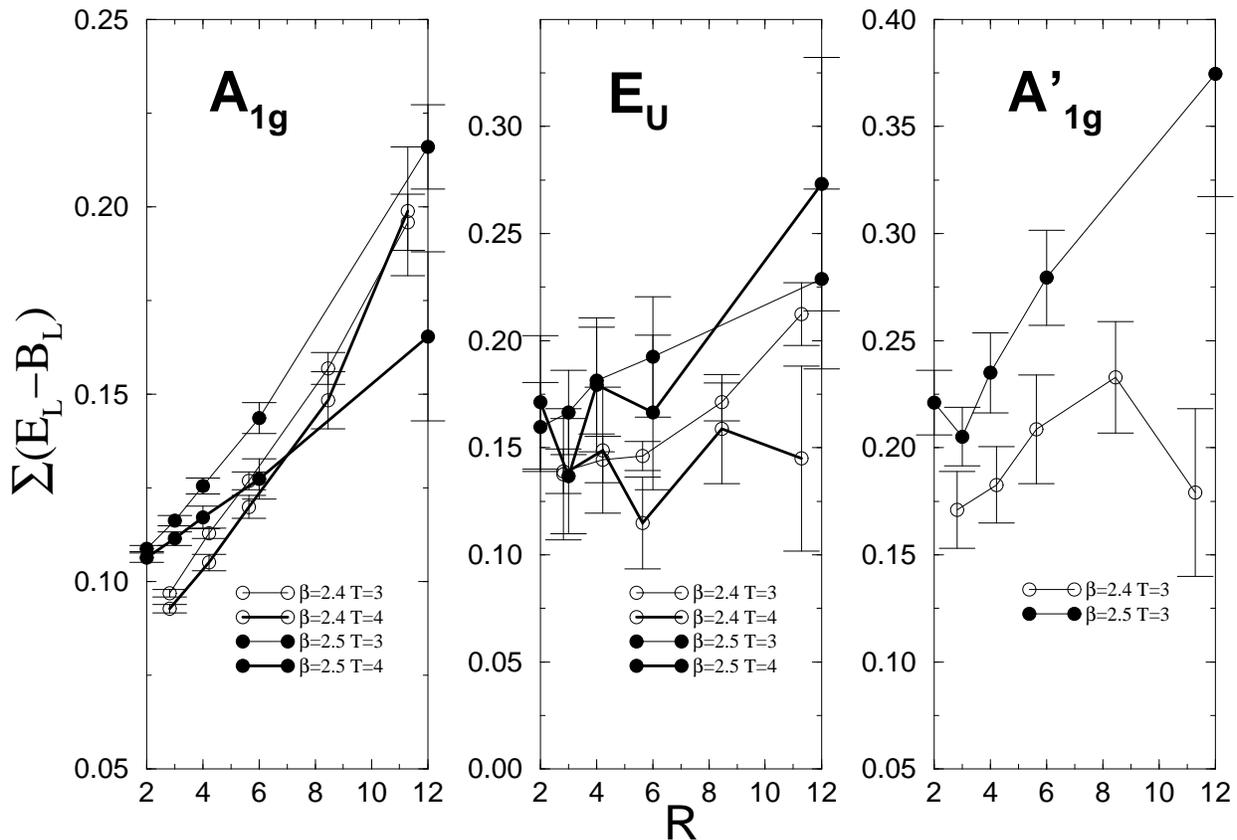}
 \caption{The scaling of longitudinal energy ($E_L$) summed over the
spatial lattice for paths with $A_{1g}$, $E_u$ and $A_{1g}'$
symmetries.}
 \label{fenls}
\end{figure}

\begin{figure}[h]
\hspace{0cm}\epsfxsize=400pt\epsfbox{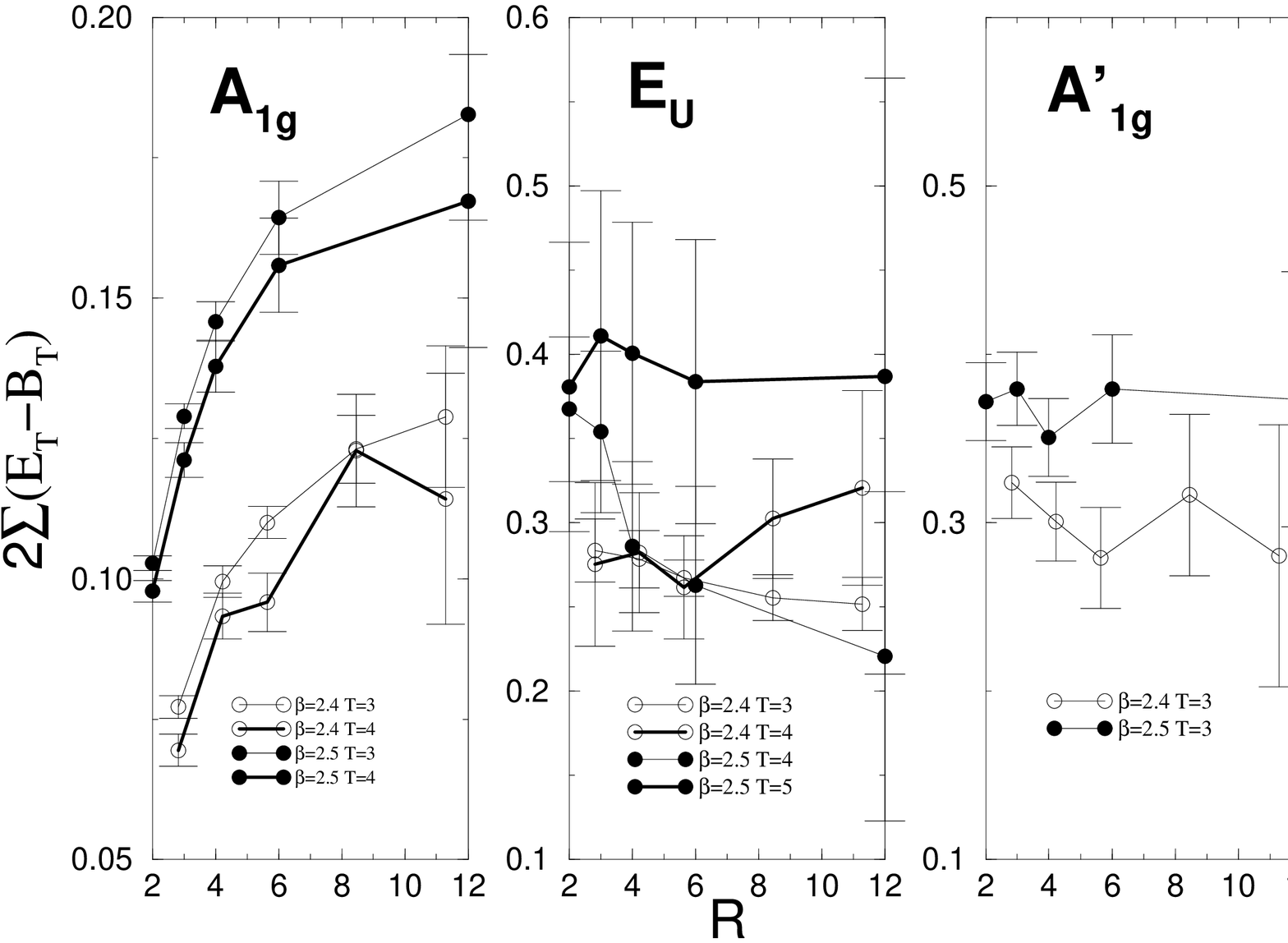}
 \caption{The scaling of transverse energy ($E_T$) summed over the
spatial lattice for paths with $A_{1g}$, $E_u$ and $A_{1g}'$
symmetries.}
 \label{fents}
\end{figure}

In Figs. \ref{facts}--\ref{fents} the three-dimensional spatial sums of
the action $S$ and longitudinal and transverse energies $E_L$, $E_T$
involved in the sum rules are  plotted as a function of $R$ and $T$ for
the flux-tube ground state  $A_{1g}$  and the two excited states $E_u$
and $A_{1g}'$ for $\beta=2.5$ and the scaled  data at $\beta=2.4$. The
basic data (${\cal E}$, ${\cal B}$) are dimensionless  and require the 
factor $\beta/a^4$ to give energy and action densities in GeV/fm$^3$.
Since  Figs. \ref{facts}--\ref{fents} show the volume integrals of the
basic data with $\beta=2.5$, for a scaling comparison, the volume
integrals of the  basic data for $\beta=2.4$ are multiplied by $2.4/(2.5
\rho)$, where $\rho$ is the ratio of lattice spacings given in Table 
\ref{tscale}.
The resulting two sets of data [$\beta=2.4({\rm scaled}),\,2.5$] should
not be expected to lie on  top of each other, since they have
self-energies and self-actions that  diverge  as $g^2/a$ from one-gluon
exchange in leading order perturbation theory.  However, the two sets of
data should be parallel to each other, since the self-energies are
independent of $R$. In the following extraction of the  beta-function,
it is the extent to which the {\em slopes} of these lines  are {\em
non-zero} that is relevant.

In Figs. \ref{facts}-\ref{fents}, for clarity,  only the data for one or
two $T$ values are  drawn.  The data for higher $T$ have larger errors,
but in most cases they are consistent with the data shown -- indicating
that the plateau in $T$ is achieved.       In Fig.~\ref{facts}, for the
action, the $A_{1g}$ and $E_u$ states show scaling and with distinct
non-zero slopes. This is best seen for the $A_{1g}$ state  and
deteriorates progressively in going to the $E_u$ and $A_{1g}'$ states. 
Scaling is questionable for the latter. Even so, the action and its
$R$-dependence are comparable in all three cases. A very approximate
estimate of the difference $\Delta S_0$ in the  $\beta=2.5$  and 
$\beta=2.4({\rm scaled})$ self-actions is given by the vertical
difference between the two sets of data. The result given by linear fits
for the  $A_{1g}$ ground state, $\Delta S_0\approx 0.12(2)$, is not
inconsistent with  the  $E_u$ case, whereas the $A_{1g}'$ excitation has
a positive  self-action difference only for the smallest $R$'s. This
curious feature of the $A_{1g}'$ data is perhaps not surprising, since
already Table \ref{th} shows that this state has considerable contamination
from other states. However, in spite of this, we thought it useful to
include such data in this paper to illustrate where the data need to be
improved.

In Fig.~\ref{fenls}, for the longitudinal energy, the data are an order
of magnitude smaller than for the action. Here the dependence on 
$\beta$ is less clear and the presence of a non-zero slope much less
distinct for the $E_u$ and  $A_{1g}'$ cases.  The difference $\Delta
E_0$ in the $\beta=2.5$  and  $\beta=2.4({\rm scaled})$ self-energies is
now best taken visually at  the lowest $R$'s, giving  $\Delta E_0\approx
0.020(5)$ for the   $A_{1g}$ state. The data for the gluonic excitations
are consistent with  this estimate. For $A_{1g}$ the curves cross at
$R=6$ and at  higher $R$'s the $\beta=2.4$ curve lies higher.

In Fig.~\ref{fents}, for the transverse
energy, only the $A_{1g}$ state shows a slope that is slightly non-zero, with 
both the $E_u$ and $A_{1g}'$ cases having slopes consistent with zero. 
However, there is a distinct self-energy effect with
$\Delta E_0\approx 0.04(1)$ for the $A_{1g}$ state, again consistent with the 
gluonic excitations.

Since it is the presence of non-zero slopes that is relevant for
extracting the beta-function, Figs. \ref{fenls} and \ref{fents} already
indicate that problems will arise   when attempting to utilize the
$E_{L,T}$ data.

\subsection{Transverse sums}

\begin{figure}[h]
\hspace{0cm}\epsfxsize=400pt\epsfbox{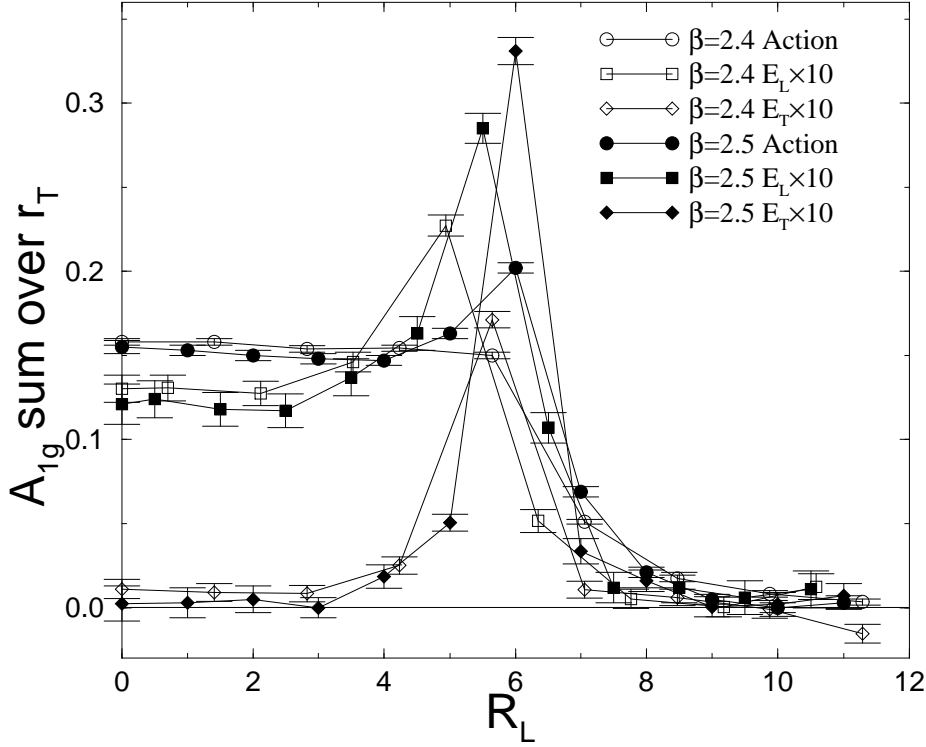}
\caption{The dependence on longitudinal position ($R_L$) of the sum  
over the transverse plane of the colour flux  contributions
corresponding to the action, longitudinal ($E_L$) and transverse  
energy ($E_T$) sum rules, eqs. \protect\ref{TASU}--\protect\ref{TEPSU}.
Here $R_L$ is measured from the mid-point for separation $R=8,12$ at 
$\beta=2.4,\,2.5$ respectively. The data are in units of $a(2.5)$ for the 
symmetric ground state (A$_{1g}$ representation) at $T=3$.}
\label{frl}
\end{figure}

\begin{figure}[h]
\hspace{0cm}\epsfxsize=400pt\epsfbox{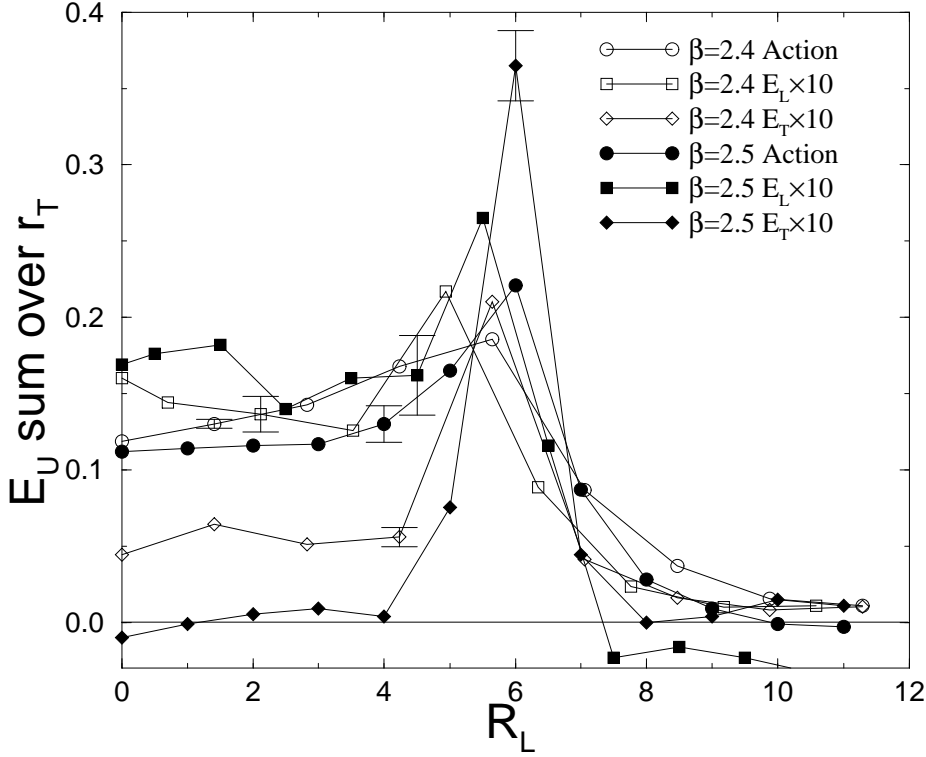}
 \caption{As in Fig.~\protect\ref{frl} but for the first gluonic
excitation ($E_u$ representation). For each data set one error bar is
shown; others are similar.}
 \label{frle}
\end{figure}

\begin{figure}[h]
\hspace{0cm}\epsfxsize=400pt\epsfbox{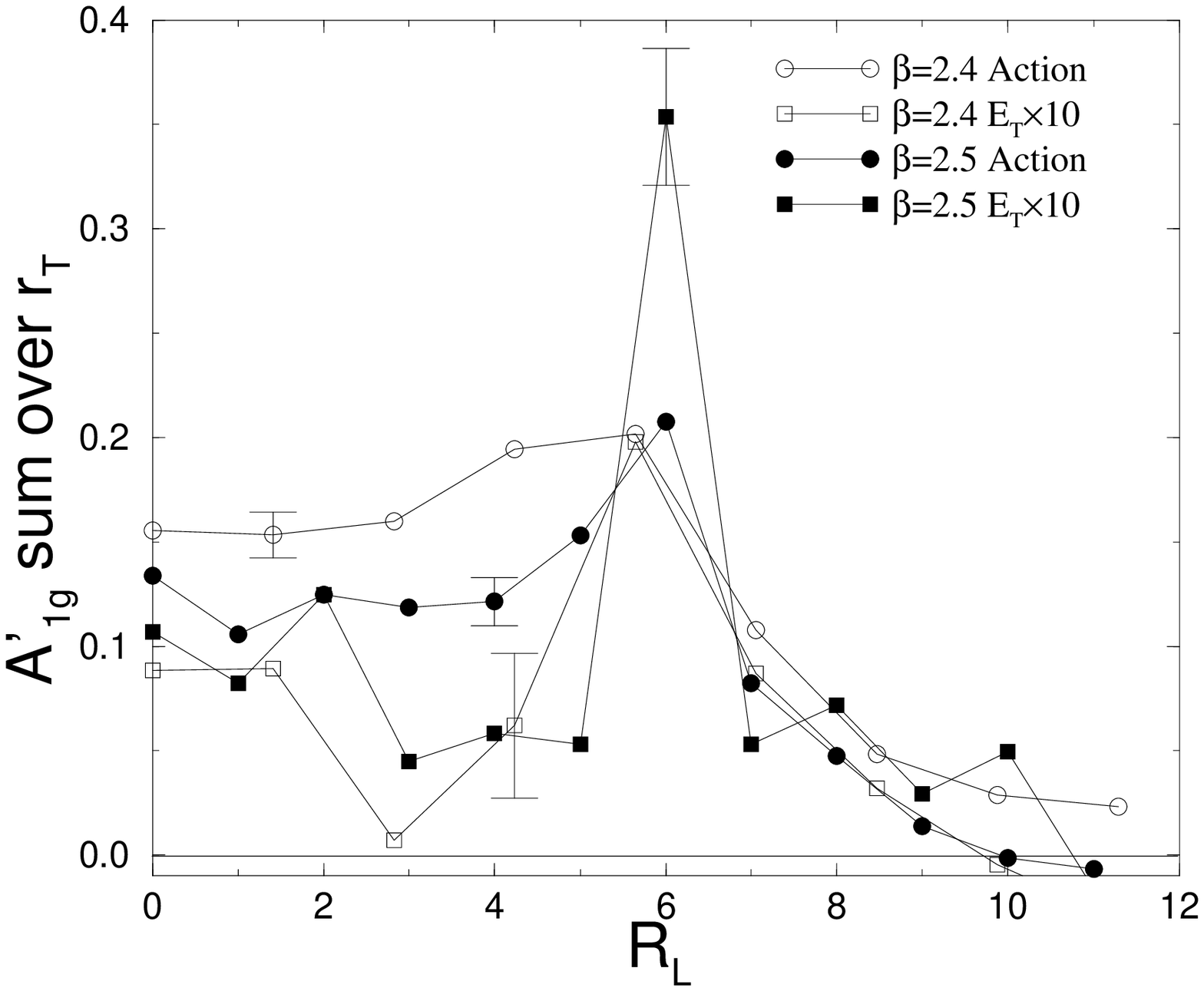}
 \caption{As in Fig.~\protect\ref{frl} but for the second gluonic
excitation ($A_{1g}'$ representation). For each data set one error bar
is shown; others are similar.} 
\label{frlp}
\end{figure}

The self-energy differences in the above spatial sums  can be seen much
more clearly in the transverse sums shown in Figs. 
\ref{frl}--\ref{frlp}, where the longitudinal dependence of the sum over
 $R_L$ is presented for paths with $A_{1g}$, $E_u$ and $A_{1g}'$
symmetries and interquark separation $R=8,\,12$ at $\beta=2.4,\,2.5$
respectively.  In these figures, $R_L=0$ corresponds to the
mid-point of the  interquark separation, while $R_L=R/2$ corresponds to
the position of the static quark sources.  The $R$  values were chosen
to correspond to approximately the same physical distance at these two
couplings, namely $0.946(4)$ and $1.007(5)$ fm.  The plotted data is
taken at $T=3$, where we have a good signal to noise ratio. 
Unfortunately this means that the excited state contamination is
relevant at $T=1,2$, where $h$ is largest as can be seen from table
\ref{th}. Furthermore, we are also using the largest $R$'s, where the
excited state  contamination  becomes quite significant especially for
the $E_u$ and $A_{1g}'$ cases.

These transverse sums do not, strictly speaking, have a continuum
limit.  However, in string models the transverse sums near the centre of
long strings should be independent of $R$. So that, to the extent that
string models are applicable and that $R$ is sufficiently large, 
scaling would be expected.  This is the assumption made in presenting 
the data in earlier works~\cite{som:88}.   In the figures, the basic
$\beta=2.5$ data is compared with the basic  $\beta=2.4$ data which has
been  multiplied by $2.4/(2.5 \rho^2)$.  The longitudinal energy is
plotted at half-integer lattice spacings (i.e. averaging over
neighbouring values of ${\cal E}_T,\, {\cal B}_L$ instead of ${\cal
E}_L,\, {\cal B}_T$) to get a better determination of the  self-energy
peak. In this case these peaks are expected to diverge  as
$g^2/a^2$ in physical units.

Fig.~\ref{frl} shows for the $A_{1g}$ state that, within error bars, all
three field  combinations (action, $E_L$, $E_T$) scale well near the
centre of the flux  tube ($R_L=0$). However, near the quarks
($R_L\approx 6$ in this case) the  $\beta=2.4$ and $2.5$ curves differ
considerably. This is mainly the effect of the self-energies being seen.
This data near $R_L=6$ shows some interesting features:

a) Except for the $\beta=2.4$ action, all the data exhibit a distinct
peak near $R_L=6$ and the $\Delta S_0, \ \Delta E_0$ extracted from
Figs. 1-3 are qualitatively consistent with the corresponding values
estimated from the areas under these peaks.  We can estimate the
self-energy peak height by subtracting the value at $R_L=0$ from the
value at $R_L\approx 6$. For the action sums at 2.4 we cannot see any
self-energy peak, while for $E_L$ the ratio  of self-energy peak heights
at $\beta=2.4$ and 2.5 is 0.59. For $E_T$ the  peak height ratio is
somewhat lower at 0.49; this will be discussed below.  In
comparison, the perturbative  expectation $(2.5 a(2.5)^2)/(2.4 a(2.4)^2)
= 0.52(1)$ is in the middle  of these values. This suggests that these
peaks are dominated by  self-energy effects. 

% X(R_L=6)-X(R_L=0):
% act 2.4 no self E
% E_L 2.4 0.09692  2.5 0.164    ratio 0.591 E0[0.1183]/E0[0.0839]=0.584
% E_T 2.4 0.160087 2.5 0.487179 ratio 0.487
%
%E_u:
% act 2.4 0.067025 2.5 0.109    ratio 0.615
% E_L 2.4 0.056898 2.5 0.096    ratio 0.592
% E_T 2.4 0.165844 2.5 0.359    ratio 0.462
%
%A_1g' next time..
%

b) The transverse energy $E_T$ is completely dominated by the
self-energy with the latter being, at least an order of magnitude larger
than  the non-self-energy terms which are expected to be essentially
independent of  $R_L$ between the two quarks. This immediately explains
the small slope of the $A_{1g}$ curve in Fig.~\ref{fents}. It also shows
that any volume integral of the self-energy  contribution cannot be
accurately evaluated on the present lattice, since --  in the 
$\beta=2.5$ case -- the $R_L=6$ contribution is an order of magnitude
larger  than those from $R_L=5,\,7$ i.e. the whole  volume integral is
given by contributions (each with $\approx 10\%$ errors)  from only {\em
three} values of $R_L$. Therefore, when the volume integrals  from
different values of $R$ are subtracted, it will be hard to get a 
meaningful signal for the $R$-dependence of interest. 

In principle, the self-energy contribution can stretch out
to the midpoint  between the two quarks. According to our data, for the
largest interquark  separations as shown here, any such contribution
seems negligible. This is seen by looking at the data near $R_L\approx
10-12$, which should be dominated by any self-energy tail. However, for
the smaller $R$'s the transverse sums at the midpoint can well have 
significant  self-energy contributions, which is seen in later in
section \ref{slr} when this data is used to determine $\beta$-functions.

c) The trend is that, for the action, the peak is about a 5\% effect compared
with the plateau contribution from $R_L=0,\ldots,6$, for $E_L$ the peak is 
about 
50\% of the plateau contribution and for $E_T$ the peak completely dominates.
Therefore, it should be expected that any predictions that depend on cancelling
the self-energies are most reliable for the action and least for $E_T$.

In Fig.~\ref{frle} the corresponding transverse sums are shown for the $E_u$ 
state. The most notable features are:

a) The action and $E_L$ both scale within error bars for $R_L\approx 0$.

b) As expected, the $\beta=2.4$ data shows, for $r_L\approx 0$, an
enhancement of $E_T$ over its $A_{1g}$ counterpart, whereas the action
and $E_L$ are comparable to  the $A_{1g}$ data. However, this
enhancement in $E_T$ is not seen for  $\beta=2.5$, but this could be due
to the relatively large errors for this  case.
     
c) Again the action for $\beta=2.4$ does not exhibit a distinct peak
near $R_L=6$. Instead it simply shows a monotonic increase as $R_L$ goes
from 0 to 6. The ratio of self-energy peak heights is 0.61, 0.59 for the
action,  $E_L$ respectively -- the latter being the same as in the
$A_{1g}$ case. For  $E_T$ the ratio is 0.46, lower than for $E_L$ by a
similar amount that was observed for $A_{1g}$. As the self-energies
should be isotropic, the difference in the peak height ratios at the two
$\beta$'s is worth exploring further.
 
 This difference is caused by the different discretisation of longitudinal and 
transverse plaquettes. 
 The electric field is dominant, so $E_T$ is mainly composed of
plaquettes lying in the transverse plane, whereas  $E_L$ is mainly  a
planar sum of plaquettes with a perpendicular orientation. A simple way
to investigate their diverging behaviour is to consider a scalar field
$\phi=e^{-b|r|}$ with $b\approx 10/{\rm fm}$ and the integrals over it 
analogous to  our sums. The (normalised) integral over the transverse
two-dimensional plane,  where the source lies, is found to change faster
when the lattice spacing is varied than the integral over a
three-dimensional ``slice'' of width $a$, analogous to $E_L$. For our 
$a(2.4)$ and  $a(2.5)$ the ratio $[{\rm Plane}(2.4)/{\rm
Plane}(2.5)]/[{\rm Slice}(2.4)/{\rm  Slice}(2.5)]\approx 0.8$ (also for
$\phi=e^{-b|r|^2/{\rm fm}}$), which agrees with the  corresponding ratio 
observed
for the $E_T$ and $E_L$ for paths with $A_{1g}$  and $E_u$ symmetries.
The lower peak height ratios for $E_T$ and the high $\sum E_T$ peaks
observed in Figs. \ref{frl}--\ref{frlp} are also due to this effect.

The diverging of the transverse sums over colour sources is 
quite consistent with the expectation from leading order perturbation theory.
$E_L$ diverges slower than the perturbative expectation, as expected,
because of the transverse extent of the dominant electric field. On the other 
hand, for $E_T$ we would expect $g^2/a^2$ behavior, while the observed peak 
height ratios diverge slightly faster than this.

d) Compared with $S$ and $E_L$, $E_T$ has a self-energy that is
comparable to or larger than the plateau contribution from $R_L=0,\ldots,6$.

In Fig.~\ref{frlp} the data for the $A_{1g}'$ state is shown with the 
following features:

a) The action and $E_T$ are approximately scaling within the rather large error
bars. However, for $E_L$ (not shown) it is not possible to make this claim 
since the error bars  are too large.

b) The $E_T$ data, unlike that in the $E_u$ case, now exhibits some
enhancement -- compared with the $A_{1g}$ state -- for {\em both} 
$\beta=$2.4 and
2.5. For example, at $\beta=2.4$ and $R_L=0$, 
$E_T(A_{1g})\approx 0.01(1), \ E_T(E_u)\approx
0.04(1)$ and $E_T(A'_{1g})\approx 0.09(5)$. Even so, the plateau terms are
still, at most, only comparable to the self-energies. Therefore, as for the
$A_{1g}$ and $E_u$ cases, those predictions that require a delicate
cancellation of the self-energies are possibly not reliable.

\subsection{Transverse profiles}

\begin{figure}[h]
\hspace{0cm}\epsfxsize=400pt\epsfbox{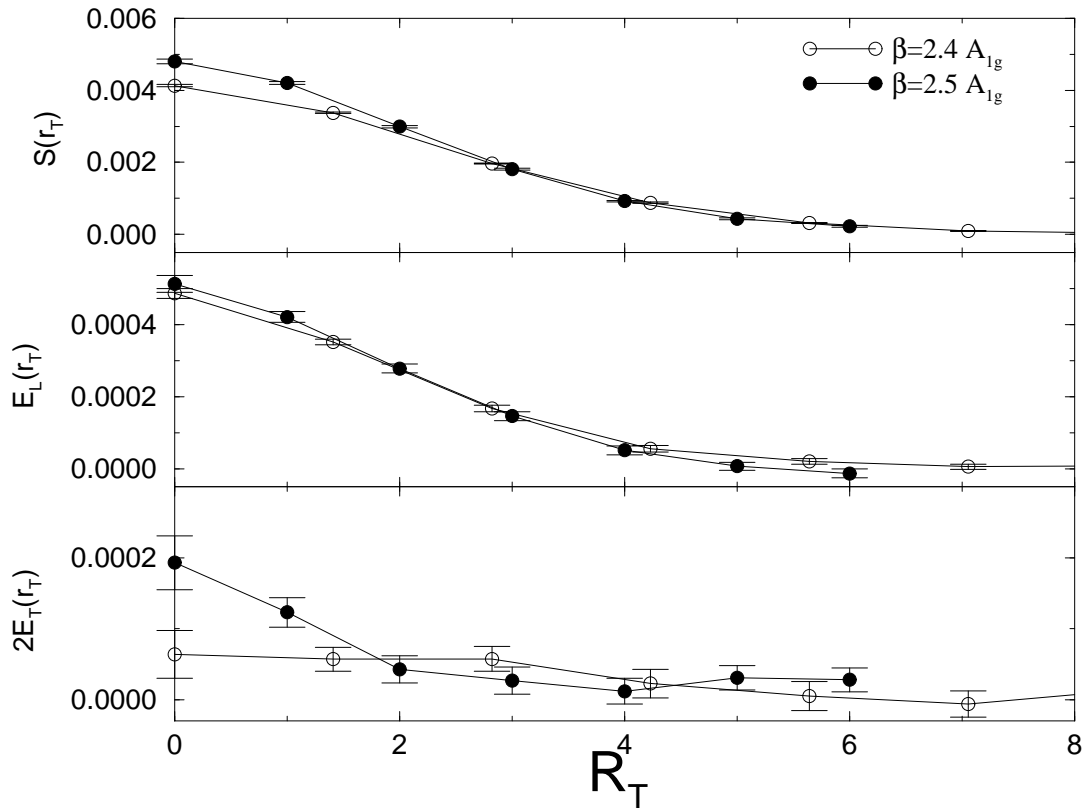}
 \caption{ The colour flux contributions corresponding to the action
($S$), longitudinal ($E_L$) and transverse energy ($E_T$) sum rules of
Eqs. \protect\ref{TASU}--\protect\ref{TEPSU} for the static quark
potential. These are shown in units of $a(2.5)$ versus transverse
distance $R_T$ at the mid-point ($R_L=R/2$) for separation $R=8,12$ at
$\beta=2.4,\,2.5$.  The data are for the symmetric  ground state
(A$_{1g}$ representation).}
 \label{frt}
\end{figure}

\begin{figure}[h]
\hspace{0cm}\epsfxsize=400pt\epsfbox{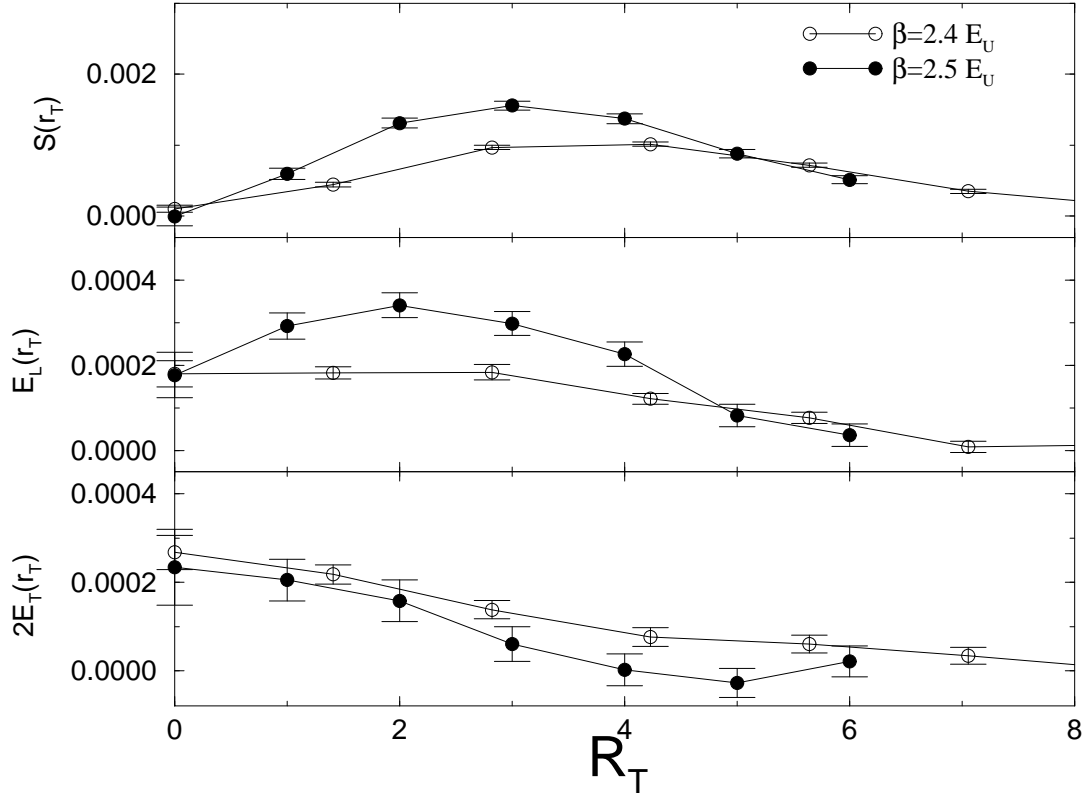}
 \caption{ As Fig.~\protect\ref{frt} but for the first gluonic
excitation ($E_u$ representation).}
 \label{frte}
\end{figure}

\begin{figure}[h]
\hspace{0cm}\epsfxsize=400pt\epsfbox{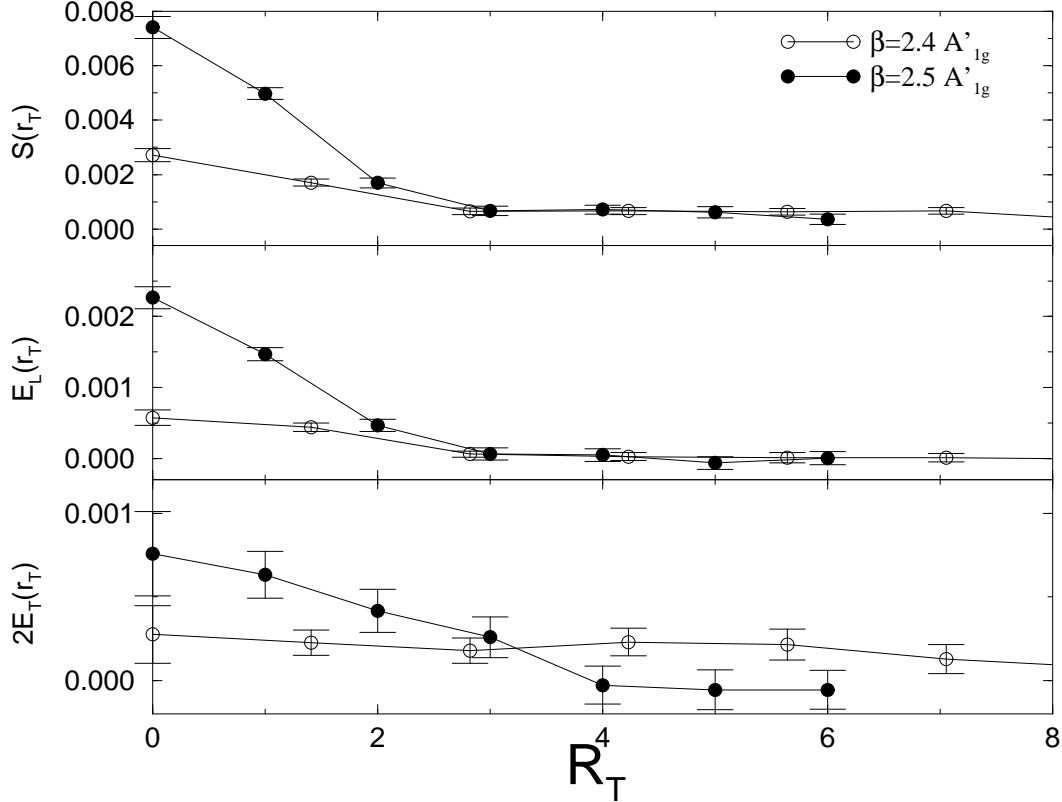}
 \caption{ As Fig.~\protect\ref{frt} but for the second gluonic
excitation ($A_{1g}'$  representation).}
 \label{frtp}
\end{figure}

In Figs. \ref{facts}--\ref{frlp} the scaling properties of the 3-d
volume and,  assuming the flux tubes have string-like features, of the
2-d transverse integrals are demonstrated -- with some combinations of
the colour fields  being more successful than others in satisfying this
property. It is, therefore, of interest to proceed finally to the
``scaling'' properties of the individual flux tube profiles. The
transverse dependence of the action ($S$),  longitudinal and transverse 
energies ($E_{L,T}$) -- measured at the midpoint ($R_L=0$ in our convention) 
with the separations $R=8,\,12$ at  $\beta=2.4,\,2.5$ respectively -- is
presented in Figs. \ref{frt}--\ref{frtp}.  The  correlations shown were
measured at $T=3$ for the Wilson loop.  As we  are again using a small
$T$ and large $R$, excited  state contamination is significant in the
$E_u$ and $A_{1g}'$ cases. Here the  $\beta=2.4$ data is compared with
the $\beta=2.5$ data by multiplying the former by $2.4/(2.5 \rho^4)$. 

Aside from any intrinsic non-scaling arising from the different
scale  of the plaquette used to probe the flux distributions, we should
also be aware  that effects can arise from the discretisation versus
$R_T$ of the distribution and,  possibly, from self-energy effects. We
have found earlier that the self-energy effects are negligible at
the midpoint ($R_L=0$) in the integrated distributions.  We here assume that 
this applies to the differential distributions so  this contribution can be
neglected. The effect of the  discretisation in $R_T$ is that a sharply 
peaked distribution will be suppressed at coarser lattice spacing. 
There is some sign  of this latter effect in our data: 
The smaller plaquette at $\beta=2.5$ should increase the observed
height of peaks (such as the  centre of the flux-tube $R_T=0$ in the
$A_{1g}$ case), whereas at the smoother  regions (away from the centre
in the $A_{1g}$ case) the shapes at the two $\beta$'s should be more
similar to each other. This is  indeed observed; in the $E_u$ case, the
largest differences are at $R_T\approx 2$ or 3 instead of $R_T=0$ as
for the $A_{1g}$ symmetry, because the distributions peak at these
values. This effect has not been mentioned in earlier works~\cite{som:88,
bal:94}.

At first sight, it appears that the only cases where the results for the two 
$\beta$'s are consistent are 
$S(A_{1g})$ and $E_L(A_{1g})$, whereas in the 2-d sums of Figs. 4-6 other
cases, such as $S(E_u), \ E_L(E_u)$ and even $S(A'_{1g}), \ E_T(A'_{1g})$
seem to show reasonable scaling. The possible reasons for this are twofold:

i) Figs. 7-9 only show the profile in a single direction -- along a
lattice axis -- whereas Figs. 4-6 are an average over all directions in
a plane.  In particular, this could have an effect on small
values of $R_T$, where rotational invariance is most violated. 

ii) The curves depicted in Figs. 7-9 must be multiplied by a phase space
factor $2\pi R_T dR_T$, when their contributions to any 2-d sum rule are
estimated. Therefore, the values near $R_T\approx 0$ get drastically
reduced and, in the $A_{1g}$ and $A_{1g}'$ cases, it is this region of
$R_T$ that is varying  the most with $\beta$.

 Assuming some function describing the continuum density we could apply
a discretisation procedure, e.g. simply averaging over cubes of
volume  $a^3$, that simulates the flattening of the peaks in our
finite-$a$  simulations. The latticized continuum function could then be
fitted  to measured points. When this procedure is applied at both
simulated values of  the lattice spacing, we would get two corrected
parameterisations. If these two agree this then would suggest that this
transverse  distribution would apply in the continuum limit.
This would be the way to compare with continuum models of the flux
tube:  indeed we find  qualitative agreement \cite{gre:96} with the
dual QCD model of Refs. \cite{bak:95,bak:96} for the $A_{1g}$
profiles.

An interesting feature in the $A_{1g}'$ profiles is a local minimum (a
dip) outside the centre of the tube as predicted by the N=1 Isgur-Paton
model  \cite{isg:85} for the energy density. In Fig.~\ref{frtp} the
action can be seen to have a  plateau at $3 \le R_T \le 5$ unlike in the
$A_{1g}$ case. For $E_L$ no evidence of a dip is found, whereas for
$E_T$ the data with  $\beta=2.4$ hints at a minimum for $R_T\approx 3$
with the $4\le R_T \le 7$  values being above zero unlike the $A_{1g}$
case, which is consistent with zero for $R_T\geq 4$.

\begin{figure}[h]
\hspace{0cm}\epsfxsize=400pt\epsfbox{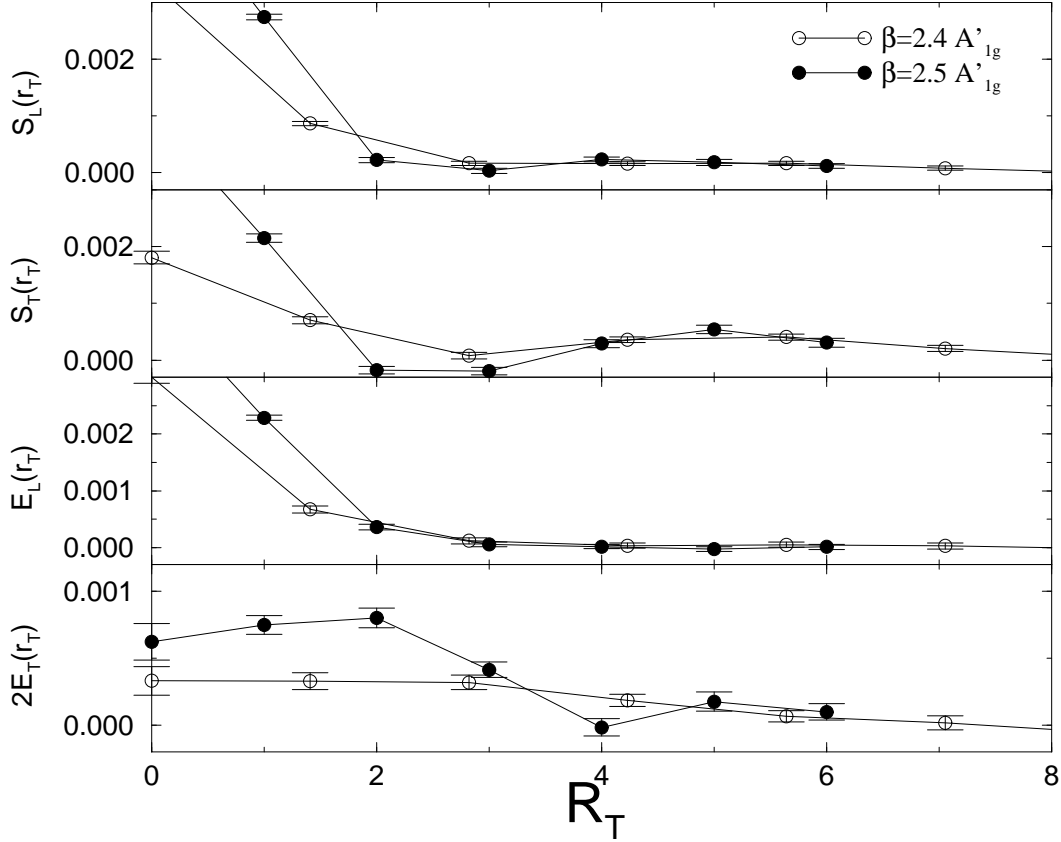}
 \caption{ As Fig.~\protect\ref{frtp} but with longitudinal and
transverse components of the action presented separately, and for
interquark separation $R=4,6$ at $\beta=2.4,2.5$ respectively.} 
 \label{frtb}
\end{figure}

A better statistical accuracy is achieved in the transverse profiles of 
flux-tubes with interquark separation $R=4,6$ at $\beta=2.4,2.5$
respectively, corresponding to interquark distances of 0.473(5) and
0.503(7) fm.  These are shown in Fig.~\ref{frtb} with the longitudinal
and transverse  components of the action plotted separately. There is
now a clear dip in the  transverse action profile, again at $R_T\approx
3$, with a corresponding maximum at  $R_T\approx 5$. A similar dip and
maximum are seen in the longitudinal action at $\beta=2.5$, whereas the
$\beta=2.4$ data has a plateau at $3\le R_T\le 6$.  As in Fig.
\ref{frtp}, no evidence for a dip is found for $E_L$, whereas the $E_T$
data shows a clear dip at $\beta=2.5$ with  a minimum at $R_T=4$. At
$\beta=2.4$ there is no minimum but the decay as a function of $R_T$ is
slow, the value at $R_T\approx 6$ being an order of magnitude higher
than in the corresponding $A_{1g}$ case.  However, again it should be
emphasized that Table 2 indicates for the $A_{1g}'$ state considerable
contamination from neighbouring states. Therefore, any nodal structure
possibly present in a pure $A_{1g}'$ state could well be smoothed out by
interference effects. Also, it should be added, that the
Isgur-Paton model which suggests such dips is less applicable for  these
smaller values of $R$. 

%**********************************************************
\section{Determination of $\beta$-functions from sum rules \label{sb}}
%**********************************************************

In Ref. \cite{mic:96}, by imposing the condition on the interquark
potential  $V$ that \\ $\partial V(R)/\partial a|_R=0$, the following
three  sum rules were derived relating $V$ to spatial sums of the
electric and  magnetic colour. 
 \begin{eqnarray} {-1 \over b} \left( V+R {\partial V \over \partial R}
\right) + S_0 & = & \sum S = -\sum  ( {\cal E}_L + 2  {\cal E}_T  + 2 
{\cal B}_T + {\cal B}_L)
 \label{TASU} \\
{1 \over 4 \beta f} \left( V+R {\partial V \over \partial R} \right) +
E_0 & = & \sum E_L = \sum  ( - {\cal E}_L + {\cal B}_L)
\label{TELSU} \\
{1 \over 4 \beta f} \left( V-R {\partial V \over \partial R} \right) +
E_0 & = & \sum E_T = \sum  ( - {\cal E}_T  +   {\cal B}_T ) \ .
\label{TEPSU} 
\end{eqnarray}
 Here the generalised  $\beta$-functions are defined considering an
asymmetric  lattice. In the notation of  Ref. \cite{mic:96}, they are
$b\equiv\partial \beta/\partial \ln a=2(S+U)$ and  $f\equiv
(U-S)/(2\beta)$. In Eqs. \ref{TASU}--\ref{TEPSU}, $S_0$ and $E_0$ are
the  self-action and -energy associated with the quarks and are,
therefore, independent of R.  The same self-energy $E_0$ is expected for
both orientations ($L$ and $T$) of  the colour electric field.

The three-loop perturbative expression for $b$ in terms of 
$\alpha=g^2/(4\pi)=1/(\pi\beta)$ is $b=-0.37151(1+
0.49193\alpha-0.9795\alpha^2+\ldots)$. On the other hand, for $f$ we
have $f=1-0.456\pi\alpha-0.25\pi\alpha b+\ldots$ where we can insert the
expression for $b$ giving $f=1-1.1408\alpha+\ldots$. 

The aim is to now extract estimates of $b$ and $f$ in the
non-perturbative situation encountered in practice. This will be carried
out in various ways, each of which has its own advantages and
disadvantages. However, their common feature is that, on the LHS of each
sum rule, the potential $V$ is measured on the lattice using Wilson
loops $W(R,T)$ and, on the RHS, as discussed in  the introduction, the
colour  fields ${\cal E}$ and ${\cal B}$ at a point $r$, are measured
using Eq.~ \ref{fmnT} involving the same loops.  Unfortunately, this
strategy is complicated by two features in Eqs. 
\ref{TASU}--\ref{TEPSU}. Firstly, the self-energies are unknown and so
their effect must be removed by considering differences between the
equations for different values of $R$ -- or for different
gluonic states $A_{1g},\, E_u,\ldots$ with the same $R$, a possibility
not considered here. Secondly, each equation  contains $\partial
V/\partial R$. Even though,  the potential $V$ itself can be readily
determined as a by-product of the  ${\cal E}$, ${\cal B}$ measurement on
a  lattice, to determine $\partial V/\partial R$ introduces some
uncertainty. All of the estimates in this section are made  using the
$A_{1g}$ ground state, where we have the best signal.  The
other gluonic states have such large errors that sensible  values of
$b,\, f$ can not be extracted. 

%*****************************************************
\subsection{Method 1: Fitting the sum rules \label{sm1}}
%*****************************************************

\begin{figure}[h]
\hspace{0cm}\epsfxsize=400pt\epsfbox{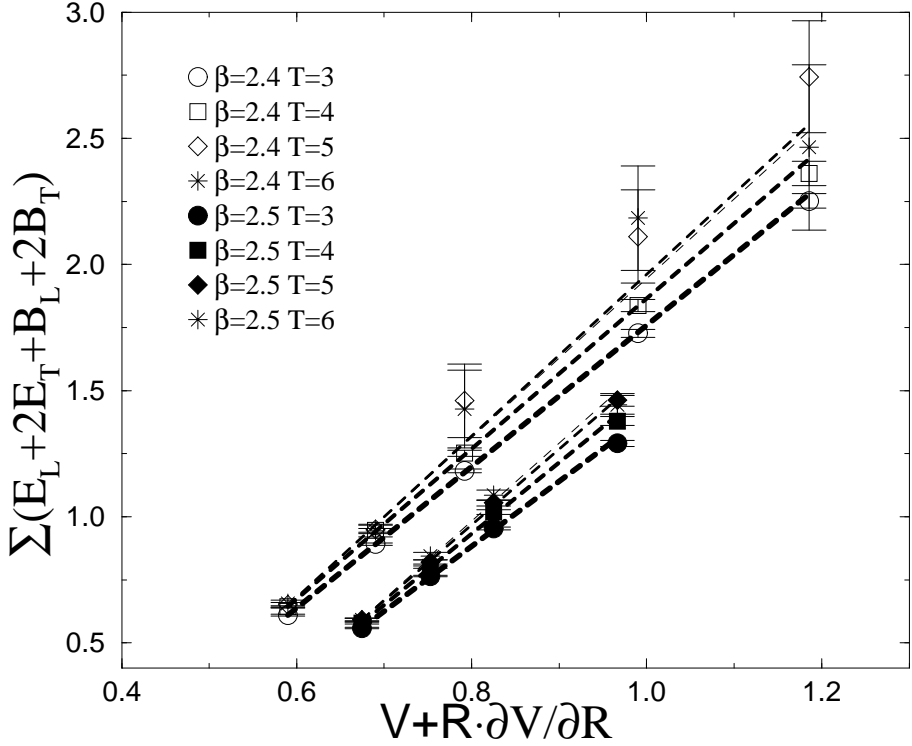}
\caption{Data corresponding to Eq.~\protect\ref{TASU} with best linear fits.}
\label{fsum1}
\end{figure}

The most direct approach is to measure ${\cal E}$ and ${\cal B}$ over
all  space and to then perform the spatial sum giving the RHS of Eqs.
\ref{TASU}--\ref{TEPSU}.  In practice, ``spatial sum'' means a sum over
a lattice that has a linear size twice that of the maximum $R$
considered i.e. for $\beta=2.4\,(2.5)$ up to $R=8\,(12)$ on lattices
with spatial size $16^3\,(24^3)$. With $V$ known numerically from  Eq.
\ref{ev}, the derivative can also be estimated. The function $b$ is then
obtained using Eq.~\ref{TASU} by plotting $\sum ({\cal E}_L+2{\cal
E}_T+{\cal B}_L+ 2{\cal B}_T)$  vs. $V(R)+R\frac{\partial V(R)}{\partial
R}$ and performing a linear fit, as  shown in Fig.~ \ref{fsum1}. The
$R=1$ points were not included in the fits due to the  artefacts they
contain, while the $R=12$ point at $\beta=2.5$  was excluded because of
its significant excited state contamination  (see Table \ref{th}). There
 are four sets of data for each $\beta$ corresponding to  the
correlation of the sum over electric and magnetic fields taken at  time
intervals $T=3,\ldots,6$. The $\beta=2.4$ data was scaled by multiplying
 with $2.4/(2.5\rho)$ to have the same units as  the $\beta=2.5$ plots.
The results of the fit in Fig.~\ref{fsum1} can be read  from the second
column of Table \ref{tbf}. 

For $\beta=2.4$ the function $b$ has reached a plateau at  $T=5$, giving
a best estimate of $-0.312(15)$, whereas for $\beta=2.5$ a  plateau has
been reached only at $T=6$ with $-0.323(9)$ being our best estimate. The
self-energy estimates $S_0$ are also seen to reach plateau at  $-1.2(1)$
and $-1.5(1)$ for $\beta=2.4,\,2.5$ respectively. 
 
Even though the $\beta=2.4$ and $2.5$ lines in Fig.~\ref{fsum1} are almost 
parallel as in Fig.~\ref{facts}, they are
separated by the difference in their self-energies. The strategy of 
fitting the data with a straight line is effectively taking differences
of Eq.~\ref{TASU} evaluated for different values of $R$ and so avoids the 
need to know $S_0$ explicitly.

In principle, the function $f$ can be extracted from either of the two sum 
rules in Eqs. \ref{TELSU} and \ref{TEPSU}. However, as shown in 
the fourth and sixth columns of Table \ref{tbf}, these predictions exhibit 
much more variation and have much greater errors than those for $b$. The main 
reason for this is due to the larger variation of the sums of the 
{\em differences} of electric and magnetic 
fields. In particular, the values of $f(I)$ at $\beta=2.4$, $T=5,6$ and 
$\beta=2.5$, $T=4,5,6$ are essentially undetermined. 

The self-energies $E_0$ are not consistent with zero for $f(I)$ at
$\beta=2.5$ as they are for $f(I,\,II)$ at $\beta=2.4$ and $f(II)$ at
$\beta=2.5$. As $E_0$ is the difference of: \\
 1) $V_0/(4\beta f)$ coming from the two-body  potentials in Eq.
\ref{VV0} and \\
 2) the self-energy in $E_L,\,E_T$ for $f(I,\,II)$ respectively, \\
 we can see that these two cancel for all cases except $f(I)$ at
$\beta=2.5$. As the self-energies should be isotropic, this is probably
caused by inaccuracies in determining $\sum E_L$. 

\begin{table}
\caption{Estimates for $b$ and $f$ at $\beta=2.4,\,2.5$ at different $T$ 
values. \label{tbf}}
\begin{center}
\begin{tabular}{l|l|c|c|c|c|c|c}
$\beta$ & $T$       & $b$       & $S_0$    & $f$ (I)  & $E_0$ & $f$ (II) & $E_0$ \\ \hline
2.4 & $3$     & -0.357(3) & -1.04(2) & 0.63(3)  & 0.004(4)  & 0.63(5) & 0.005(3) \\
    & $4$     & -0.336(5) & -1.11(3) & 0.71(6)  & 0.010(7) & 0.74(10) & 0.007(4) \\
    & $5$     & -0.312(15) & -1.24(10) & 1.5(1.3)  & 0.06(2) & 0.84(28) & 0.01(1) \\
    & $6$     & -0.317(21) & -1.21(14) & 4.5(7.0)  & 0.09(4) & 1.4(8) & 0.02(2) \\ \hline
2.5 & $3$     & -0.389(4) & -1.17(2) & 0.89(7)  & 0.033(7)  & 0.59(3)  & 0.002(3)  \\
    & $4$     & -0.354(5) & -1.32(3) & 1.41(23) & 0.059(10)  & 0.64(5)  & 0.003(5)  \\
    & $5$     & -0.333(7) & -1.44(4) & 1.49(38) & 0.063(16) & 0.75(11) & 0.011(8) \\
    & $6$     & -0.323(9) & -1.49(7) & 6(11) & 0.10(3) & 0.74(16) & 0.009(12) \\
\end{tabular}
\end{center}
\end{table}

\begin{figure}[h]
\hspace{0cm}\epsfxsize=400pt\epsfbox{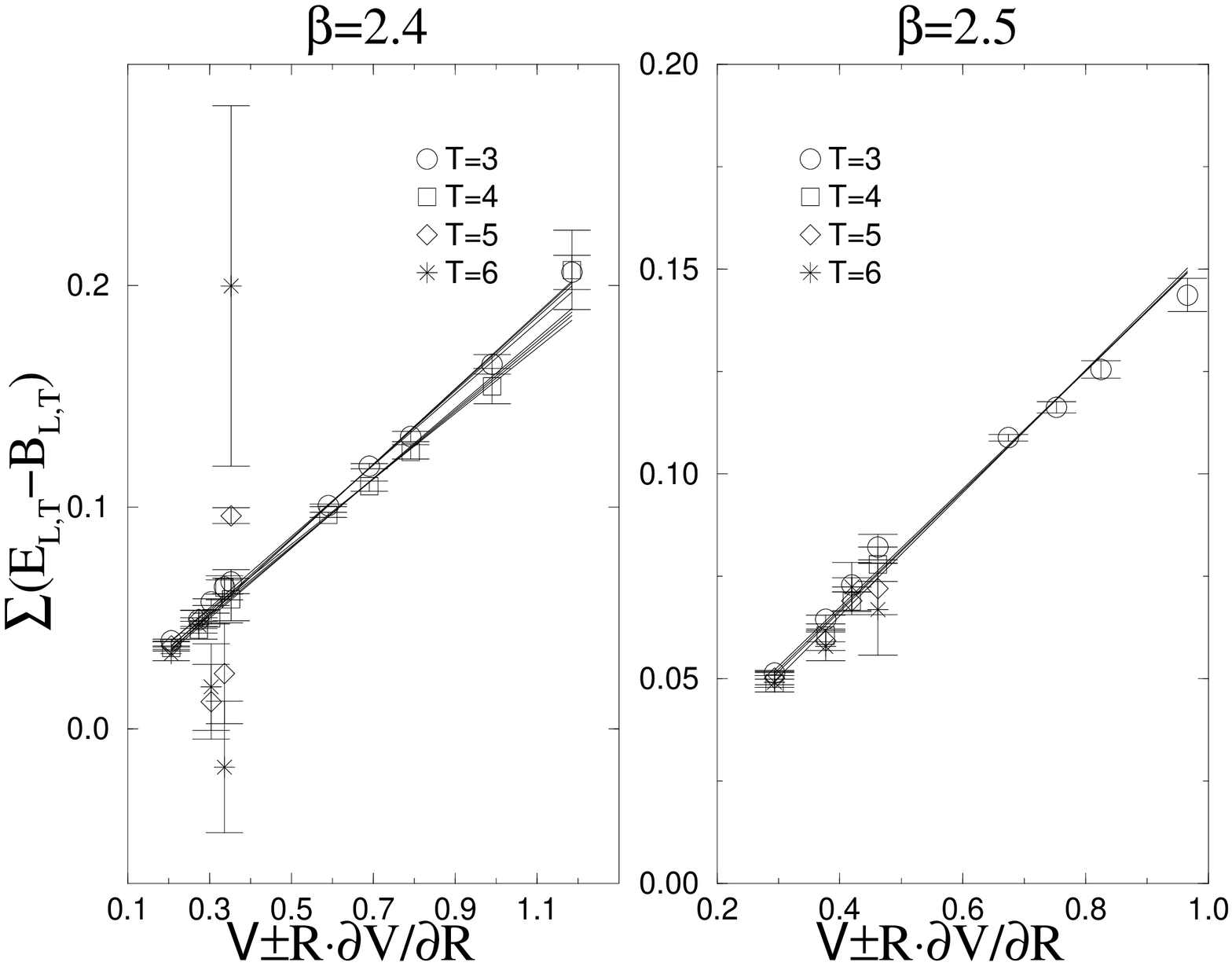}
\caption{Determination of $f(I+II)$ by combining two sum rules.}
\label{f25}
\end{figure}

This inaccuracy in determining $f$ can be greatly reduced if the sum
rules in eqs. \ref{TELSU} and \ref{TEPSU} are fitted together. At
$\beta=2.4,\,2.5$ data for the former sum rule is taken at $T=3,4;\,T=3$
 respectively, while for the latter it can be taken at any $T$ from 3 to
6.  The function $f$ obtained in this manner is presented  in Fig.
\ref{f25} and Table \ref{tf} and can be seen to lead to a much more 
accurate estimate of $f$. Our  best estimates are 0.65(1) and 0.68(1) at
$\beta=2.4, 2.5$ respectively. Also the values of self-energy
are now more stable at  $E_0 \approx 0.01(1)$ -- a number that is about
two orders of magnitude smaller than the self-actions $S_0$.

\begin{table}
\caption{Combined fits of $f$ at $\beta=2.4,\,2.5$ at different $T$ values. 
The first $T$ value refers to the data used for Eq.~\protect\ref{TELSU}, the 
other to the data used for Eq.~\protect\ref{TEPSU}. \label{tf}}
\begin{center}
\begin{tabular}{l|l|c|c}
$\beta$ & $T$    & $f(I+II)$ & $E_0$ \\ \hline
2.4     & $3,3$  & 0.647(7)  & 0.006(1) \\
        & $3,4$  & 0.616(8)  &  0.0008(10) \\
        & $3,5$  & 0.627(17) & 0.003(3) \\
        & $3,6$  & 0.612(18) & 0.013(4) \\
        & $4,3$  & 0.705(13) & 0.009(1) \\
        & $4,4$  & 0.671(13) & 0.004(1) \\
        & $4,5$  & 0.684(23) & 0.040(5) \\
        & $4,6$  & 0.659(27) & 0.030(6) \\ \hline
2.5     & $3,3$  & 0.694(11) & 0.010(1)  \\
        & $3,4$  & 0.667(11) & 0.005(1)  \\
        & $3,5$  & 0.682(16) & 0.008(2) \\
        & $3,6$  & 0.688(21) & 0.009(3) \\
\end{tabular}
\end{center}
\end{table}

In the above, the parameter $\rho$ has been extracted and found to be
1.411(13). This then suggests as a direct estimate of $b$, averaged over the
$\beta$ range of 2.4 to 2.5, the value
\[b=\frac{\Delta \beta}{\Delta\ln[a]}=
 -\frac{2.5-2.4}{\ln[a(2.5)]-\ln[a(2.4)]}=-0.290(8).\]

Even though this is admittedly a very crude estimate, it should represent the
average value of $b$ over this range of $\beta$. However, it appears to be  
slightly
smaller in magnitude than the average of the above estimates $-0.312(15)$ at 
$\beta(2.4)$ and $-0.323(9)$ at $\beta (2.5)$.
The origin of this two sigma difference is not clear.

%*********************************************
\subsection{Method 2: Combining the sum rules}
%********************************************

\begin{figure}[h]
\hspace{0cm}\epsfxsize=400pt\epsfbox{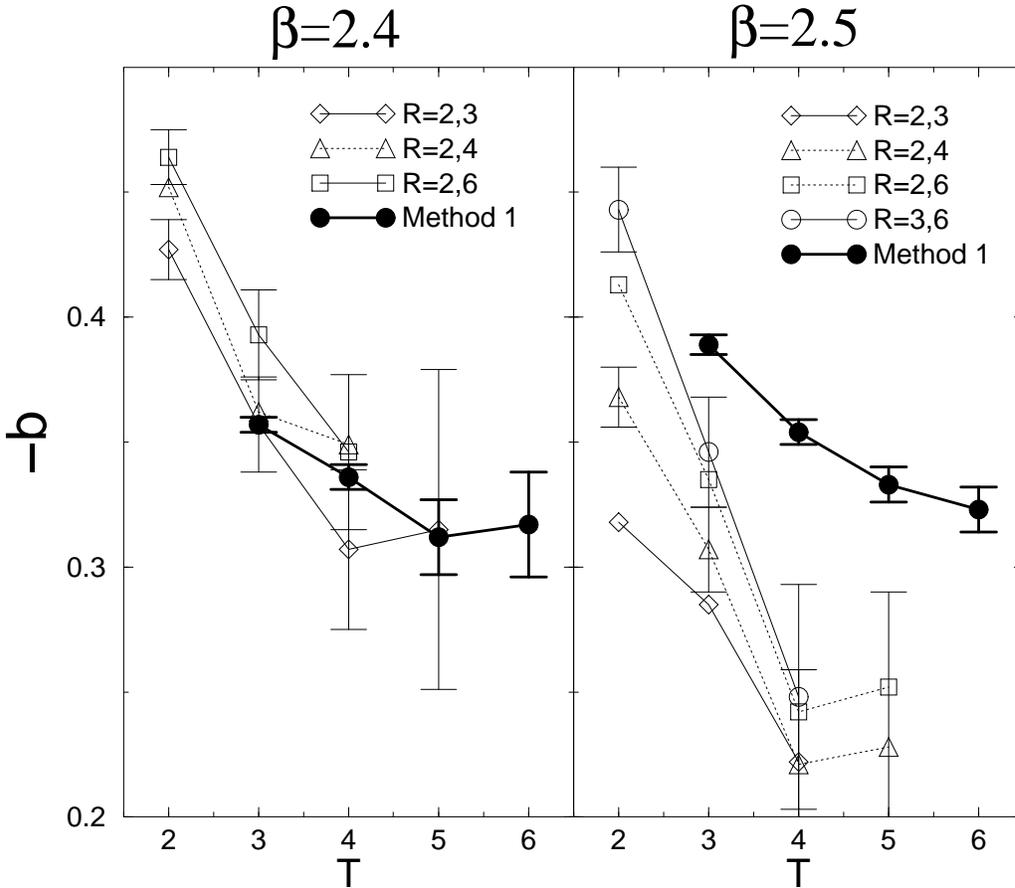}
 \caption{Estimates of $b$ from Eq.~\protect\ref{eb} for a) $\beta=2.4$
b) $\beta=2.5$.}
 \label{fb}
\end{figure}

As shown in Ref. \cite{mic:96b}, 
one way to avoid estimating $\partial V(R)/\partial R$ and $S_0$, $E_0$ is to 
explicitly eliminate them by writing down Eqs. \ref{TASU}-\ref{TEPSU} for two 
different values of $R$. In this way
\begin{eqnarray}
b & = & \frac{ 2 ( V(R_1)-V(R_2)) \left( 1+\frac{
 \sum  (E_T)_{R_1}  -
 \sum  (E_T)_{R_2} }{
 \sum (E_L)_{R_1}-
 \sum (E_L)_{R_2}
}
\right)^{-1}
}{
 \sum  S_{R_1}
 -\sum S_{R_2}
} \label{eb}\\
f & = & \frac{V(R_1)-V(R_2)}{2\beta [\sum (E_T)_{R_1}-
 \sum (E_T)_{R_2}+ 
\sum (E_L)_{R_1}-
\sum (E_L)_{R_2}
]} \label{ef} .
\end{eqnarray}
 At first sight this appears to be what is needed - expressions that
involve quantities that can be measured directly. However, in practice,
there is a problem -- $b$ becomes dependent on the differences $\sum
(E_L)_{R_1}-\sum (E_L)_{R_2}$ and  $\sum (E_T)_{R_1}-\sum (E_T)_{R_2}$
from Eqs. \ref{TELSU}--\ref{TEPSU}, and, as seen above, the values of
these differences are less accurate  than $\sum S_{R_1} - \sum S_{R_2}$.
The outcome of  this strategy is given in  Fig.~\ref{fb}. There it is
seen that the $b(2.4)$ results are consistent with  those given by
method 1 -- but have much larger error bars. However, the $b(2.5)$
results are essentially inconsistent with method 1. A similar problem
arises  with the values of $f$ from Eq.~\ref{ef}. Again $f(2.4)$ is
consistent with the earlier estimates of method 1 in Ref. \cite{mic:96b}
-- but with much larger error bars. For example, with  $R_1,R_2=2,6$ we
get $f(T=4,5)= 0.64(7),0.64(15)$. However, compared with method 1,
$f(2.5)$ is badly out -- rising to $\approx 0.9(2)$ at $T\approx 4,5$.
It should be added that this is not a problem of the measurements being 
poorly distributed, since plotting the bootstrap values  of $b$ shows
that the errors are not  underestimates due to asymmetric non-gaussian
bootstrap distributions.

This difference between the two estimates of $b$ for $\beta=2.5$ but not
2.4 is caused by the inaccurate determination  of the longitudinal
energy sums $\sum (E_L)_{R_1}-\sum (E_L)_{R_2}$ in  Eq.~\ref{eb}.  As
can be seen in Fig.~\ref{fenls}, for the $\sum E_L$ curves  at
$\beta=2.5$ the slope is rather erratic and indicates a  much smaller
value than at $\beta=2.4$. This is also reflected in  Table \ref{tbf},
where the  $f(I)$ value obtained using the $\sum E_L$ values is already
at $T=3$  unrealistically high for $\beta=2.5$, getting worse with
increasing $T$. This means that the $\sum (E_L)_{R_1}-\sum (E_L)_{R_2}$
are underestimated. For $\sum E_T$ the situation is more consistent,
which can be seen in the larger slope in Fig.~\ref{fents} and the
reasonable  behavior for the $f(II)$ in Table \ref{tbf} at $\beta=2.5$.
Thus  $\sum (E_L)_{R_1}-\sum (E_L)_{R_2}$ is too small and  $\sum
(E_T)_{R_1}-\sum (E_T)_{R_2}$ realistic, leading to underestimates of
$b(2.5)$ using Eq.~\ref{eb} and an overestimate of $f(2.5)$ from Eq.
\ref{ef}.  The signal being worse for $E_L$ than $E_T$ is somewhat
surprising, since it is $E_T$ where the self-energy completely
dominates. Therefore,  one would have expected it to be harder to get a
signal  for $E_T$, because it requires  a more delicate cancellation of
the  self-energy.

Why doesn't this happen at $\beta=2.4$? From Figs. \ref{fenls} and 
\ref{fents} we can see that the $\sum E_L$ slope is larger and the 
$\sum E_T$ slope slightly smaller than at $\beta=2.5$. This is again
reflected in  Table \ref{tbf}, where $f(I)$ again gets larger with
increasing $T$,  but not as much as for $\beta=2.5$ and with larger
errors making the  estimates consistent with a realistic value. At the
same time, unlike for $\beta=2.5$, $f(II)$  also increases. This
fortuitously leads to a realistic ratio of the  differences  of the sums
in Eq.~\ref{eb} and a value of $b$ consistent with method 1. 

\subsection{Method 3: The large $R$ limit \label{slr}}

In the above, the derivatives of $V$ are calculated numerically from the
lattice form of the interquark potential in Eq.~\ref{ev}. However, for 
sufficiently large $R$ ($R\geq 2$), the continuum form of $V$ (i.e. with
$[1/R]_L$  replaced with $1/R$) is a good approximation. When, in
addition to this, the effect of the self-energies is  removed by
evaluating the sum rules at two values of $R$, Eqs.
\ref{TASU}--\ref{TEPSU} reduce to
 \begin{eqnarray}
 b & = & \frac{-2b_S(R_1-R_2)}{\sum S_{R_1} - \sum S_{R_2}} \label{elrb}\\
 f(I) & = & \frac{b_S(R_1-R_2)}{2\beta [\sum (E_L)_{R_1} - \sum (E_L)_{R_2}]} \label{elrf1}\\
 f(II) & = & \frac{-e(1/R_1-1/R_2)}{2\beta [\sum (E_T)_{R_1} - \sum (E_T)_{R_2}]}. \label{elrf2}
\end{eqnarray}

\begin{figure}[h]
\hspace{0cm}\epsfxsize=400pt\epsfbox{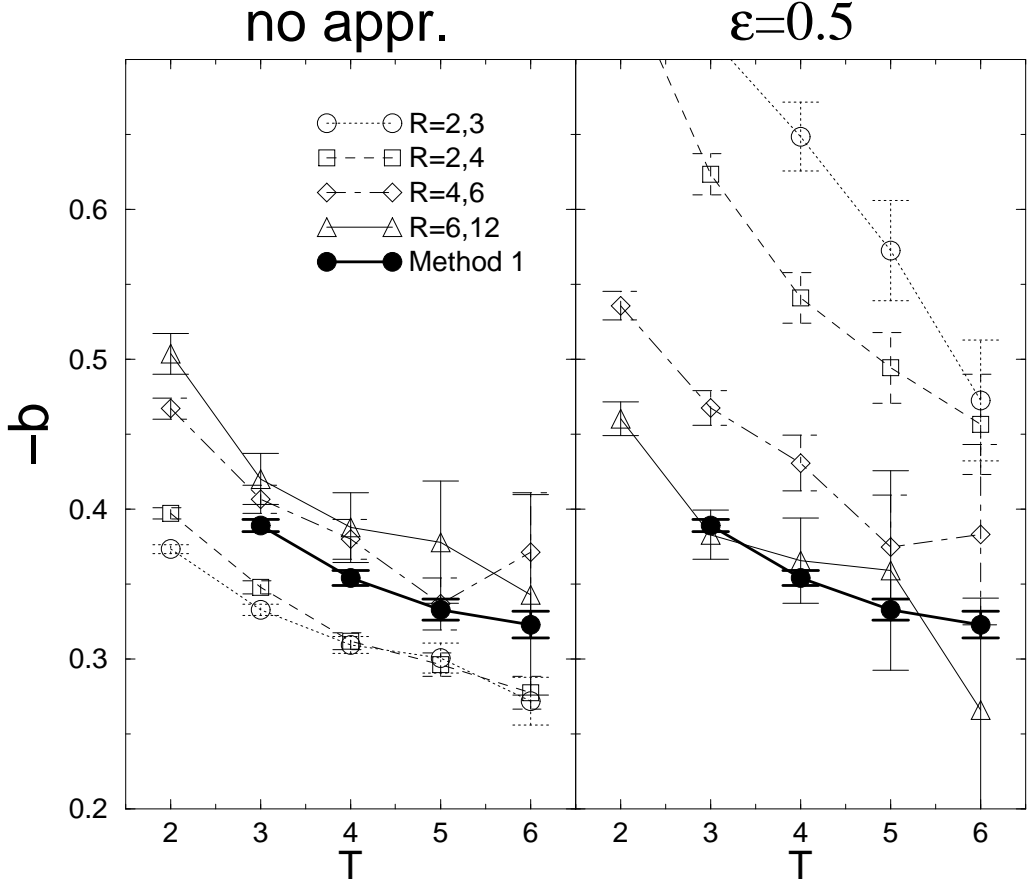}
 \caption{Estimates of $b$ at $\beta=2.5$ using a) the continuum form of
$V(R)$ (Eq.~\ref{elrb}) and, additionally, b) a constant longitudinal
profile approximation for the flux tube with $\epsilon=0.5$ (Eq.
\protect\ref{elrp}.)}
 \label{fmethod3}
\end{figure}

As seen in figure \ref{fmethod3} a) for $\beta=2.5$,  Eq.~\ref{elrb}
gives estimates in agreement with the fits in  Table \ref{tbf} for
$R_1,R_2=2,\ldots,12$. However, this is not surprising, since there the
results are an average over a range of $R$ values, whereas Eqs.
\ref{elrb}--\ref{elrf2} can be considered as an average using simply two
 values of $R$. A similar situation holds for $\beta=2.4$. 

Of more interest is the large $R$ limit, where it is expected from
string models that both the action and energy flux-tubes should have a
form that is essentially constant for all $R_L$ in the range $-R/2\le
R_L\le R/2$ with $R$-independent self-action or -energy terms
concentrated at $R_L\approx\pm R/2$. As can be seen in Fig.~\ref{frl},
the sums over the transverse plane  in the middle of the quarks agree at
$\beta=2.4$ and 2.5, supporting the accuracy of our estimates in this
region. Therefore, the action difference in Eq.~\ref{elrb}, 
 $\Delta \sum S= (\sum S_{R_1}-\sum S_{R_2})$ should be well approximated by \\
$\Delta \sum S= [R_1S_{R_1}(R_L=0)-R_2S_{R_2}(R_L=0)]$  and similarly for
$\Delta \sum E_L$ and $\Delta \sum E_T$ in Eqs. \ref{elrf1} and \ref{elrf2}. 

In practice, since $R_{1,2}$ are not very large, it is probably more
realistic to include a correction for the \lq{}overshoot\rq{} of the plateau
term at  $R_L\approx \pm R/2$ i.e.
 \begin{equation}
\Delta \sum S \rightarrow \Delta \sum S'=
[R_1({\rm eff})S_{R_1}(R_L=0)-R_2({\rm eff})S_{R_2}(R_L=0)],
 \label{elrp} 
 \end{equation} 
 where $R_i({\rm eff})=R_i+2\epsilon$. Here $\epsilon$ is the
\lq{}overshoot\rq{} at  each end of the plateau. From Figs.
\ref{frl}--\ref{frlp}, it is seen that  $\epsilon$ must be approximately
one half of a lattice spacing.

The results at $\beta=2.5$ with $\epsilon=0.5$ are shown in 
Fig.~\ref{fmethod3} b). The values of $b$ approach those given by method 1 
as $R_{1,2}$ increases, agreeing when $R_{1,2}=6,12$. The 
disagreement at the smaller $R$'s is mainly due to significant 
self-energy contributions at the midpoint, which don't cancel as they vary 
with $R$. The good results from $R=6$ thus show that the self-energy 
contributions at midpoint are negligible at these interquark separations.
The same approximation 
can be used in the spatial sums of method 1.
With $R=2,3$ excluded from the fit of Eq.~\ref{TASU} we get 
$b=-0.35(4)$ at $T=6$ and $\beta=2.5$. 

The extracted value of $b$ is only  weakly dependent on $\epsilon$ even
for the rather small values of $R_{1,2}$ used here. This situation will
only improve as  $R_{1,2}$ increases. In  principle, this method has two
advantages over the earlier ways of extracting $b$ and $f$:

1) It avoids the need for any delicate numerical cancellations of
self-energies, since they are assumed to have canceled exactly in the 
$\Delta$'s.

2) Only the single 2-dimensional sum over the $R_L=0$ plane is
necessary. This avoids the full 3-dimensional sum of 
Eqs. \ref{TASU} --\ref{TEPSU}. This
estimate can be further improved by averaging the 2-dimensional integrals over
the smallest values of $R_L$. However, in practice this would not lead to 
savings in 
computer time, since averaging over all positions and orientations of 
the Wilson loops would, in any case, mean using all possible planes in the 
lattice.

It has one disadvantage, however, in that it assumes a string-like 
longitudinal dependence of the colour flux distribution.

For large $R$ one would expect that the string tension is given
by the longitudinal energy density in the transverse plane at the midpoint:
$$
b_S = 2\beta f E_L(R_L=0).
$$
Taking the values of $E_L(R_L=0)$ from Fig.~\ref{frl} and using our best 
estimates of $f$ gives $\sqrt{b_S(2.4)}=483(15),\, \sqrt{b_S(2.5)}=477(24)$ 
MeV. These are close to the values given by fits to experimental spectra. 

%********************************************
\subsection{Comparison with other approaches}
%********************************************

There are two main ways to extract $\beta$-functions in lattice gauge 
theories. First, one can measure observables at different  values of
some parameter (such as a coupling or a quark mass) and then try to
estimate the response of the observable to a change in that parameter.
This can be carried out using  either finite differences or an
interpolating function. Both of these introduce a systematic error
either from the use of a  finite interval or  from the choice of the
interpolating function. In the second method, one  can use the fact that
response functions  are related to correlation functions via lattice sum
rules. This method has  less systematic errors but often leads to large
statistical errors, because   the observables involve delicate
cancellations. 

In the case of SU(2) gauge theory, the first method has been used
recently in  Ref. \cite{eng:95}, where $b$ was estimated by measuring
the critical  temperature at six values of the coupling and using an
exponential ansatz or a spline interpolation for the $\beta$ dependence
of $a$. In the same work, non-perturbative derivatives of couplings in
the time and space directions with respect to the asymmetry factor
$\xi=a_s/a_t$ were estimated from $b$ and measurement of the free
energy. In Ref.  \cite{bal:94} lattice spacings were  determined from
the string tension at seven couplings in the range  $2.3\le \beta \le
2.85$ and fitted with an  ansatz of a linear $a$ dependence of
$\Lambda_{\rm lat}$. The errors on these  estimates do not include 
systematic effects from the use of a linear approximation.  In Ref. 
\cite{pen:96b} the lattice spacing was determined  at six couplings
ranging from $2.3\le \beta \le 2.55$ through Sommer's  equation  $r_0^2
F(r_0) = c$ discussed after Eq.~\ref{ev}. Fitting of the three-loop 
relationship between $a$ and  $\beta$, with extra terms and systematic
errors estimated using different  values of $c$ and forms of the fitted
function, lead to results consistent with Ref. \cite{eng:95}. 

The sum rule method was first explored in Ref. \cite{bal:95}, where 
$\partial V(R)/\partial \beta$ was fitted from the time  dependence of
Eq.~\ref{fmnT} and $b$ then extracted through a second fit of  $\partial
V(R)/\partial \beta$ vs $R$. Optimistically also potentials at  $R=1$
were included in the fits. As this was only a feasibility study
the estimate of $b=-0.25(2)$ at $\beta=2.5$ should be considered as 
preliminary.  A direct formula for $b$, Eq.~\ref{eb}, involving
differences of two-body  potentials at two values of $R$ and their
action sums was derived in Ref.  \cite{mic:96b}. An  average over pairs
of $R$ values gave the estimate $b=-0.35(2)$ at  $\beta=2.4$. The values
at some $R$ pairs were outside this estimate and  doubts about the
convergence in $T$ remained, as the correlations were  calculated only
up to $T=4$ for larger $R$'s. 

Table \ref{tb} collects the most reliable estimates of $b$ at
$\beta=2.4$ and 2.5, whereas the available estimates of $f$ are shown in
Table \ref{tcf}. Here it should be noted that an effective coupling
larger than the bare coupling moves the perturbative estimates of $f$
closer to the  non-perturbative values, whereas the opposite happens for
$b$. A perturbative evaluation of $b$ is thus unreliable. The non-perturbative
estimates of $b$ agree with each other, except for the one from 
Ref.~\cite{bal:94}, which has a significant systematic error. Averaging the 
three consistent estimates gives $b(2.4)=-0.306(6),\, b(2.5)=-0.316(4)$, which
are 78\%, 81\%, respectively, of the three-loop predictions. For $f$ our
estimates are not far from those in Ref.~\cite{eng:95}, with similar
ratios of the non-perturbative and perturbative estimates as for $b$.

\begin{table}[htb]
\begin{center}
\begin{tabular}{l|c|c|c|c|c} 
$\beta$  & Our estimate & Ref. \cite{eng:95} & Ref. \cite{pen:96b} & Ref. \cite{bal:94} & 3-loop PT \\ \hline
2.4      & -0.312(15) & -0.3018  & -0.305(6)   & -0.330(4)  & -0.3893 \\
2.5      & -0.323(9)  & -0.3115  & -0.312(2)   & -0.340(4)  & -0.3889 \\ 
\end{tabular}
\caption{Comparison between values of $b \equiv \partial \beta/\partial \ln a$
\label{tb}}
\end{center}
\end{table}

\begin{table}[htb]
\begin{center}
\begin{tabular}{l|c|c|c|c|c} 
$\beta$  & Our estimate & Ref. \cite{eng:95} & PT \\ \hline
2.4      & 0.65(1)  & 0.72  & 0.85 \\
2.5      & 0.68(1)  & 0.74  & 0.86 \\ 
\end{tabular}
\caption{Comparison between values of $f \equiv (U-S)/(2\beta)$
\label{tcf}}
\end{center}
\end{table}

%********************
\section{Conclusions} 
%********************

In this paper the spatial distribution and nature  of the electric
and magnetic colour fields  between two quarks are measured for
$\beta=2.4$ and $2.5$. We discuss carefully the theoretical 
expectations for scaling versus $\beta$ of these distributions and
compare with our  results.  For the  observables with a well-controlled
continuum limit (three-dimensional sums), scaling is investigated and found
to be good in most cases of interest. For more differential  observables
(transverse sums and -profiles) the changes seen  between the two 
values of $\beta$ can be explained qualitatively from the
discretisation:   the plateaux stay the same, while the peaks get higher
for smaller $a$. Self-energies, as measured by transverse plane sums,
diverge  in a manner suggested by leading order perturbation theory,
with the  divergence being faster in the $E_L$ case because of the
longitudinal extent of the plaquettes measuring the  dominant electric
field. By comparing various  combinations of the spatial sums of the
${\cal E},\,{\cal B}$ fields with the  interquark  potential $V(R)\pm
\partial V(R)/\partial R$, estimates can be made of the  generalised
$\beta$-functions $b$ and $f$ -- see Eqs. \ref{TASU}--\ref{TEPSU}. 

There are two problems that prevent the direct use of these equations. 
Firstly, since the whole calculation is performed on a lattice, $V(R)$
is only known at discrete values of $R$, so that the values of $\partial
V(R)/ \partial R$ can not be measured directly. Secondly, both $V(R)$
and  the spatial sums have self-action or -energy contributions. To
minimize or  avoid these problems the extraction of $b$ and $f$ can be
carried out in  different ways each of  which has its own advantages and
disadvantages concerning statistical and systematical errors. 

The most direct approach -- method 1 of subsection \ref{sm1} -- is to 
parameterize $V(R)$ so that $\partial V(R)/\partial R$ can be readily 
calculated. This overcomes the $\partial V(R)/\partial R$ problem at the
expense of some systematic error introduced by the form of
parameterisation of  $V(R)$ used. By plotting, as a function of $R$, 
$V(R)\pm \partial V(R)/\partial R$ versus the various spatial sums, the
slopes give immediately $b$ and $f$. This resulted in
$b(2.4)=-0.312(15)$  and $b(2.5)=-0.323(9)$. However, the value of $f$
was much more uncertain when the two sum rules -- one involving $\sum
E_L$ and the other $\sum E_T$ -- were used separately. When these two
sum rules were fitted simultaneously, a better result emerged, $f(2.4) =
0.65(1)$ and $f(2.5)=0.68(1)$. Already at  this stage, it could be seen
that the data causing most of these uncertainties are those
involving the longitudinal energy $\sum E_L$ especially for 
$\beta=2.5$.

One way to avoid the $\partial V(R)/\partial R$ and self-action and
-energy problems is to combine the original three sum rules in such a
way  as to eliminate the problems -- see Eqs. \ref{eb}--\ref{ef} for
method 2. However, in spite of the ideal form of these equations they
result in  poor estimates of both $b$ and $f$. For $\beta=2.4$, $b$ and
$f$ agree with method 1 but with large statistical errors -- see Fig.
\ref{fb} a). But for  $\beta=2.5$ the values of $b$ tend to be too low
and of $f$ too high -- both  with large error bars. These differences
can be directly attributed to the inaccuracy  of the data for $\sum
E_L$.  

The third method for extracting $b$ and $f$ exploits the form 
expected of  $V(R)$ and the flux tube profiles in the limit
$R\rightarrow\infty$ from  string models.  When a continuum
parameterisation of $V(R)$ is used and two $R$ values are subtracted,
the sum rules reduce to those in Eqs. \ref{elrb}--\ref{elrf2} and these
yield results consistent with those of method 1, but with  larger
statistical errors. In the large $R$ limit, a further approximation is
to replace the flux tube profile by one that is constant between the
quarks ($-\frac{R}{2} \le R_L \le \frac{R}{2}$) and dropping to zero
rapidly for $R_L\ge\frac{R}{2}$ and $R_L\le -\frac{R}{2}$. At the
largest  $R$'s the results are consistent with method 1, but with large
statistical errors. These results are only very weakly dependent on the
point where the profile drops  to zero beyond the positions of the
quarks. 

As seen in tables \ref{tb} and \ref{tcf}, our best estimates for $b$ and
$f$  agree with other recent non-perturbative estimates, most
importantly the  finite temperature approach of Ref. \cite{eng:95}. It
thus seems safe to  conclude that order $a^2$ effects in the
extraction of the $\beta$-function are small at the $\beta$-values
studied using the methods described.  Thus we have a unique
non-perturbative $\beta$-function which describes the  deviations from
asymptotic scaling at these values of the coupling. 

For the state with $A_{1g}'$ symmetry, our data shows the existence of a
dip in the action and transverse energy distributions away from the
centre of the flux-tube in the transverse plane between the quarks. This
qualitatively confirms the prediction of the  Isgur-Paton flux-tube
model for the energy distribution. No such node is seen  for the
longitudinal energy, and for the action it is more pronounced in the 
transverse component. 

In the future we plan to study four-body flux distributions and 
their relationship to two-body flux-tubes. Sum rules will be used verify 
the correctness of the measured distributions. 

\section{Acknowledgements}

The authors wish to thank P. Spencer for discussions during the initial 
stages of this work. Funding from the 
Finnish Academy (P.P.) is gratefully acknowledged. Most of the simulations
were performed on the Cray C94 at the CSC in Helsinki.

\end{document}